\documentclass[secnumarabic,twocolumn,graphics,floatfix,nofootinbib,
tightenlines,nobibnotes,aps,prb]{revtex4-1}

\usepackage{subfigure}
\usepackage{color,amsmath,graphicx,epsfig,amssymb,picture}
\usepackage[T1]{fontenc}
\usepackage{pslatex}
\usepackage{txfonts}

\usepackage[svgnames]{xcolor}
\usepackage{hyperref}
\hypersetup{colorlinks,breaklinks,
            urlcolor=Blue,
            linkcolor=Blue,citecolor=Blue}

\newcommand{\vv}[1]{\mathbf{#1}}

\begin{document}

\def\pd#1#2{\frac{\partial #1}{\partial #2}}
\def\bk{{\bf k}}
\def\bx{{\bf x}}
\def\br{{\bf r}}
\def\d{{\text d}}
\def\grad{{\mbox{\boldmath$\nabla$\unboldmath}}}

\title{\bf Vortex Nucleation Limited Mobility of Free Electron Bubbles in the\\ Gross-Pitaevskii Model of a Superfluid}
\author {Alberto Villois and Hayder Salman}
\affiliation {School of Mathematics, University of East Anglia, Norwich Research
Park, Norwich, NR4 7TJ, UK}

\begin {abstract}
We study the motion of an electron bubble in the zero temperature limit where neither phonons nor rotons provide a significant contribution to the drag exerted on an ion moving within the superfluid. By using the Gross-Clark model, in which a Gross-Pitaevskii equation for the superfluid wavefunction is coupled to a Schr\"{o}dinger equation for the electron wavefunction, we study how vortex nucleation affects the measured drift velocity of the ion. We use parameters that give realistic values of the ratio of the radius of the bubble with respect to the healing length in superfluid $^4$He at a pressure of one bar. By performing fully 3D spatio-temporal simulations of the superfluid coupled to an electron, that is modelled within an adiabatic approximation and moving under the influence of an applied electric field, we are able to recover the key dynamics of the ion-vortex interactions that arise and the subsequent ion-vortex complexes that can form. Using the numerically computed drift velocity of the ion as a function of the applied electric field, we determine the vortex-nucleation limited mobility of the ion to recover values in reasonable agreement with measured data. 
\end{abstract}
%\pacs{67.25.dk, 47.37.+q}
\maketitle

\section{Introduction}
Electrically charged particles have been one of the most effective probes to study properties of liquid helium in the superfluid state. Beginning with the pioneering works of Williams \cite{Williams1957}, Careri et al. \cite{Careri1959} and Reif and Mayer \cite{Reif1960}, it has been observed that ions moving through liquid helium due to an external applied electric field can interact with different types of excitations that act to produce a drag force on the ion\cite{Borghesani2007}. On the one hand, upon exceeding a critical velocity, these ions can nucleate vortex rings. On the other hand, phonons and rotons scattering off the ion also provide an important contribution to the drag force experienced by the ion. These characteristics allow ions to be used to study microscopic hydrodynamic structures that form when a critical velocity is exceeded. At the same time, they provide useful probes to glean information regarding the properties of a quantum turbulent flow.

In this work we will mainly focus on the study of so called \emph{electron-bubbles}. The existence of electrons in the so-called self-trapped bubble state was initially suggested to explain their anomalous low mobilities of negative ions that were measured at low temperatures in superfluid $^4$He \cite{Careri1959,Ferrel1957,Kuper1961}. The rationale behind this model is that it is energetically favourable for a single electron to carve out a spherical cavity within the superfluid due to the short-range repulsive interactions that would otherwise exist between the bare electron and the cloud of electrons of the helium atoms. Although the electron in the self-trapped bubble state has received further experimental confirmation \cite{Classen1998}, there are many aspects characterising the dynamics of these ions that remain obscure. In particular, the detailed dynamical mechanisms that give rise to the drag forces acting on electron bubbles at low pressure and high electric fields remains poorly understood \cite{Nancolas1985}. 

Difficulties in directly observing the relevant microscopic hydrodynamic structures has meant that many of the proposals that have been put forward to explain observed measurements have not been fully verified. At the same time, direct numerical modelling of the problem has been hindered by the lack of an accurate microscopic model that can be used to study the complex spatio-temporal dynamics of the ion interacting with the superfluid. Relatively recently, there has been some work employing density-functional theories\cite{Ancilotto2010,Jin2010,Jin2010JLTP,Aitken2016}, that can accurately reproduce the equation of state (as well as the roton dispersion relation), in order to study the dynamics of the electron bubble. However, given the complexity of these models, simulations were restricted to axisymmetric configurations which we believe to be inadequate in representing some of the key physics such as the mechanism of asymmetric capture of the ion by nucleated vortex rings. Motivated by these questions and possibilities that electron bubbles provide in measuring properties of quantum turbulence in the zero temperature limit \cite{Walmsley2007,Walmsley2008}, we will aim to uncover the dynamics of electron bubbles by focussing on the key hydrodynamic processes that determine the limiting velocity of the ion as a function of an applied electric field.

Since there is no universally accepted microscopic model for liquid helium, we will adopt the so-called Gross-Clark model\cite{Gross1958,Clark1965} to study the 3D motion of an electron bubble within a superfluid. In this model, a Schr\"{o}dinger equation describing the wavefunction of the electron is coupled to a mean-field equation of a superfluid. In this work, we will adopt a Gross-Pitaevskii (GP) equation for the superfluid. We note that such a model does not provide an accurate description for $^4$He since it neither reproduces the correct equation of state nor does it describe the correct dispersion relation since a roton minimum is not present. However it has been shown by Berloff and Roberts \cite{Berloff2001} that this model can account for the deformations affecting the bubble in its motion and it also captures all the main qualitative physics characterising the interaction between electron bubbles and superfluid vortices. We note that it has recently been shown that a vortex filament description of a superfluid can be systematically derived from the GP equation\cite{Bustamante2015}. Therefore, despite the shortcomings of the GP  model in accurately representing certain properties of superfluid $^4$He, we anticipate that the model is reasonably accurate in allowing us to infer the hydrodynamic interactions of quantised vortices with the negative ion impurity.

\section{Mathematical Model}
\subsection{The Gross-Clark Model}\label{Sec:GrossClarkModel}
We begin by adopting the Gross-Clark model\cite{Gross1958,Clark1965} in which superfluid $^4$He is modelled by a GP equation.
The energy of the system is then given by the Hamiltonian
\begin{align}\label{Eq:GC_Htot}
H=H_{GP}+H_{e}+H_{GP-e} \, .
\end{align}
Here liquid helium is governed by the GP Hamiltonian
\begin{align}\label{Eq:GPHAM}
H_{GP}=\int \left(\frac{\hbar^2}{2m_4} |\nabla \psi |^2 +\frac{V_0}{2}|\psi|^4\right)\d^3\vv{x},
\end{align}
where $m_4$ is mass of the $^4$He atom, whereas the electron is represented by
\begin{align}\label{Eq:H_SCH}
H_{e}=\int\left(\frac{\hbar^2}{2m_e} |\nabla \phi |^2  +eQy|\phi|^2\right)\d^3\vv{x}\, ,
\end{align} 
where $m_e$ is mass of the electron. In order to study the transport of the ion through the liquid, we have included the second term which models the effect of an applied constant electric field $Q$ directed along the $y$-coordinate direction of the domain and $e=1.6\times 10^{-19}C $ is the electric charge of the electron.
We model the interaction between the superfluid and the electron by the term
\begin{align}
H_{GP-e}=\int U_0|\psi|^2|\phi|^2\d^3\vv{x} \label{Eq:H_GP_e}\, .
\end{align}
In this model the parameters $U_0=2\pi l \hbar^2/m_e$ 
and $V_0=4\pi d \hbar^2/m_4$
represent the two-body short-range fermion-boson and the boson-boson interactions with effective scattering lengths given by $d$ and $l$, respectively. Variation of $H$ with respect to $\psi^*$ and $\phi^*$ results in the equations of motion
\begin{align}
i\hbar \frac{\partial \psi}{\partial t}& =-\frac{\hbar^2}{2m_4}\nabla^2 \psi+( U_0 |\phi|^2 +V_0 |\psi|^2)\psi \, , \label{Eq:GP_GC}\\
i\hbar \frac{\partial \phi}{\partial t}&=-\frac{\hbar^2}{2m_e}\nabla^2 \phi+U_0 |\psi|^2\phi +eQy\phi\, . \label{Eq:SCH_GC}
\end{align}
The wavefunctions are subject to the normalization conditions
\begin{align}%\label{eq:normpsi}
\int |\psi|^2 \d^3\mathbf{x}=N \, , \,\,\,\,\, \text{and} \,\,\,\,\,
\int |\phi|^2\d^3\mathbf{x}=1 \, ,
\label{Eq:normphi}
\end{align}
where $N$ denotes the total number of $^4$He atoms. The GP equation provides the simplest model capable of reproducing the key phenomena characterising the interaction between an ion and quantum vortices. For these purposes, it is essential to ensure that the model that recovers the correct ratio of the radius of the ion relative to the healing length. As previously discussed in [\onlinecite{Berloff2000,*Berloff2001}], the GP model contains sufficient parameters that allows the model to be tuned to recover this property. On the other hand, the compressibility of the fluid will be represented inaccurately. In fact, in the GP equation, the pressure of the liquid is given by 
\begin{align}\label{Eq:P_GP}
p=\frac{V_0|\psi|^4}{2} \, ,
\end{align}
which provides an inaccurate relation between pressure and density for liquid $^4$He. Although other models have been proposed that remedy this deficiency of the GP equation\cite{Berloff2009,*Jin2010}, in this work we are interested in regimes where the motion of the ion is strongly dominated by the presence of superfluid vortices. Therefore, provided phonon emission is the not the dominant contribution to the drag which is expected to the case for experiments at low temperatures and low pressures, we can expect this to be less important than accurate modelling of the interaction of vortices with the ion. Similarly, the lack of a roton in the dispersion relation is of less concern since in the low pressure and low temperature regimes, the density of roton excitations diminishes very rapidly. Moreover, experimental measurements indicate that they play a less important role in comparison to the process of vortex ring nucleation which is believed to be the main contributing factor to the drag exerted on the ion for sufficiently high electric fields\cite{Borghesani2007}.

\subsection{Non-dimensional Form of the Equations of Motion \label{Sec:Nondim}}
In order to gain further insight into the properties of the electron in the self-trapped bubble state and to identify the key length scales that will arise in our problem which need to be well resolved within our numerical simulations, we will adopt a simple model of a perfectly spherical cavity at equilibrium. Assuming that the electron is in its $s$-state and is trapped within a cavity of radius $b$. For simplicity, the cavity is assumed to have infinite depth. It can then be shown (see Appendix {\ref{App:Cavity}}) that the total energy for the electron bubble-superfluid system is then given by
\begin{align}\label{eq:Energy_Bubble1}
E=E_q+E_V+E_T=\frac{\hbar^2\pi^2}{2 m_e b^2}+\frac{4\pi b^3}{3}p+4\pi Tb^2.
\end{align}
The first contribution to the total energy corresponds to the quantum mechanical energy associated with the zero-point motion of the electron. The second contribution is determined by the work required to carve out a cavity within the superfluid due to the pressure field $p=V_0 \rho^2/2m_4^2$ for a spherical cavity. The third contribution to the total energy of the system is proportional to the area of the bubble and it can be associated to the surface tension, $T$, of the cavity wall. %Details of how to derive Eq.~\eqref{eq:Energy_Bubble1} from the equations presented in \S~\ref{Sec:GrossClarkModel} is given in the Appendix.

Using this model, we can now estimate the radius $b$ of the electron bubble and subsequently its hydrodynamic mass $m_{h}$ \cite{Batchelor1967}. Since the electron mass $m_e$ is much smaller than the mass of the $^4$He atom, $m_4$, with $\delta={m_e}/{m_4}\sim1.4\times 10^{-4}$,
%\begin{equation}
%\delta=\frac{m_e}{m_4}\sim1.4\times 10^{-4},
%\end{align}
the effective mass of the bubble $(m_e+m_h)$ can then be approximated by its hydrodynamic mass which is given by
\begin{align}\label{Eq:EffM}
m_h=\frac{2}{3}\pi \rho b^3.
\end{align}
At zero pressure, Eq.~\eqref{eq:Energy_Bubble1} can be used to evaluate the radius of the bubble that minimizes the electron energy $E$; this gives
\begin{align}
b=\left(\frac{\pi\hbar^2}{8m_e T} \right)^{1/4} \, .
\end{align}
Using typical measured values of parameters for liquid helium at zero temperature, such as the surface tension of bulk helium\cite{Roche1997}, $T=375 \, \mu \text{J m}^{−2}$, and the liquid density $\rho= 0.145\, \text{g/cm}^3$, we can finally estimate that the effective radius is $b=18.91 \text{\AA}$ whereas the mass $m_h=309\,m_4$ for an electron bubble at zero pressure.\\

For non-zero pressure, it is possible to estimate the radius of the bubble by using the method of dominant balance (see Appendix \ref{App:Cavity}) under the condition that $\delta\rightarrow 0$. The respective radius of the bubble is then given by
\begin{align}
b \sim \left( \frac{\pi \hbar^2}{4m_e p}\right)^{1/5} \, .
\label{Eq:radius_b}
\end{align}
The radius, $b$, provides an important length scale in the problem that dictates the size of the computational domain that will be needed in our simulation to resolve the relevant physical scales of interest.\\

Having identified the typical radius of the bubble, we can now integrate the superfluid-electron system numerically by rewriting the equations of motion in non-dimensional form.  
We begin by introducing the transformations
\begin{align}\label{Eq:Transf}
\vv{x}\rightarrow a\vv{x} ,\quad t\rightarrow \sigma t ,\quad \psi\rightarrow \Psi_{\infty}\psi ,\quad \phi\rightarrow \Phi \phi  ,\quad Q \rightarrow q Q \, ,
\end{align}
such that positions are measured in units of the superfluid healing length given by
\begin{align}
a=\frac{\hbar}{\sqrt{2m_4 \mu}}=(8\pi d \Psi_{\infty}^2)^{-1/2} \, ,
\end{align}
where $\mu$ denotes the chemical potential for a uniform condensate wavefunction $\Psi_{\infty}$ with $N$ particles, i.e.\ $\Psi_{\infty}=\sqrt{{\rho_{\infty}}/{m_4}}=\sqrt{{\mu}/{V_0}}$. The time-scale is set by the healing length, $a$, and the speed of sound, $c$, such that
\begin{align}
\sigma=\frac{a}{\sqrt{2}c}=\frac{\hbar}{2\mu} \, .
\end{align}
%\begin{align}
%\Psi=\sqrt{\frac{\rho_{\infty}}{m_4}}=\sqrt{\frac{\mu}{V_0}}.
%\end{align}
Using the re-scalings given by Eq.~\eqref{Eq:Transf}, Eq.~\eqref{Eq:GP_GC} transforms to
\begin{equation}
i \frac{\partial \psi}{\partial t} =-\frac{1}{2}\nabla^2 \psi+\frac{1}{2}\left( 4\pi a^2\left[\frac{m_4 l }{m_e a }\right]\Phi^2|\phi|^2 + |\psi|^2\right)\psi.
\end{equation}
We, therefore, introduce the small parameter
\begin{align}
\epsilon=\left( \frac{a m_e }{l m_4}\right)^{1/5} \, .
\end{align}
Noting that $1/\epsilon$ is of the same order as the dimensionless radius of the bubble $b/a$, we chose to rescale the electron wave function such that  
\begin{align}
\Phi=\left(\frac{\epsilon^3}{4\pi a^3}\right)^{1/2} \, , \text{with} \,\,\,\,\,
\int |\phi|^2 \d^3\vv{x}=\frac{4\pi}{\epsilon^3} \, .
\end{align}
Finally, we express the electric field in units of
\begin{align}
q=\left(\frac{\mu}{\delta e a  }\right) \, .
\end{align}

The above rescaling allows us to rewrite Eqs.~\eqref{Eq:GP_GC} and \eqref{Eq:SCH_GC} as
\begin{align}\label{GP_ad}
i \frac{\partial \psi}{\partial t} &=-\frac{1}{2}\nabla^2 \psi+\frac{\gamma}{2}|\phi|^2\psi +\frac{1}{2}|\psi|^2\psi \, , \\
\label{SCH_ad}
i\frac{\partial \phi}{\partial t} &=-\frac{1}{2\delta}\nabla ^2\phi +\frac{1}{2\delta}\left(\zeta^2|\psi|^2+yQ\right)\phi \, ,
\end{align}
where
\begin{align}
\gamma=\frac{1}{\epsilon^2} \qquad \zeta^2=\frac{l}{2d} \qquad \delta=\frac{m_e}{m_4} \, .
\end{align}
Motivated by modelling ions in superfluid $^4$He, we follow [\onlinecite{Berloff2000PRB,*Berloff2001}] and take $a=1\, \text{\AA} $, $\zeta=0.41$, $\epsilon=0.187$, $\mu=5.22\times10^{-4} \, \text{eV}$, $\delta=1.4\times 10^{-4}$ and $\rho_{\infty}=0.145 \, \text{Kg/cm}^2$. This gives the unit of the electric field  
$q=3.72 \, \text{V/\AA}$, the unit of time $\sigma=0.63\times 10^{-12} \, \text{s}$, and the unit of velocity ${a}/{\sigma}={\sqrt 2} c = 1.58\times 10^2 \, \text{m/s}$.

\subsection{Adiabatic approximation \label{Sec:Adiabatic}}
The non-dimensional form of the equations presented above reveals a major difficulty arising from any attempt to directly integrate these equations using realistic values of parameters for superfluid $^4$He. In particular, the small value of $\delta$ appearing in the Schr$\ddot{\text{o}}$dinger equation~\eqref{SCH_ad} leads to a clear disparity in the time scales of the superfluid and the electron.
Therefore time resolved solutions of Eqs.~\eqref{GP_ad} and~\eqref{SCH_ad} for scenarios of physical relevance becomes impractical. Although the disparity in time scales leads to numerical challenges, one can also exploit this inherent feature of the system in order to
eliminate the source of the difficulty. In particular, we observe that, for an electron trapped within the potential $|\psi(\vv{x},t)|^2$ created by the surrounding fluid, if the time scale over which the potential changes is much larger than the typical quantum time scale $m_e b^2/\pi \hbar$ of the electron (set by Eq.~\eqref{Eq:KinEn_Sphere}), then we are in a regime where the so called adiabatic (also known as Born-Oppenheimer) approximation holds. In quantum mechanics, the \emph{adiabatic theorem} states that for adiabatic changes of the potential that do not lead to degenerate eigenmodes, a particle initially in the $n$'th-eigenstate, $\phi_n$, at time $t_i$, will remain in that $n$'th-eigenstate $\tilde{\phi}_n$ at time, $t_f$, but will acquire some extra phase factors, such that the final state is given by   
\begin{align}
\tilde{\phi}(t_f)=\phi_n(t_i) e^{i\theta_n(t_f-t_0)}e^{i\chi_n(t_f-t_0)} 
\end{align}
where $\theta_n$ and $\chi_n$ are called the dynamical and the geometrical phase factors, respectively \cite{Messiah1961}.

We remark that the condition on the degeneracy of eigenmodes, and consequently the validity of the adiabatic approximation, can breakdown during the splitting of an electron bubble. This scenario can occur when an electron bubble that contains an electron in an excited $p$-state is subjected to a negative pressure pulse that can cause the bubble to split into two parts\cite{Maris2000,Wei2015}. Under such situations, the adiabatic approximation is no longer applicable since the splitting of the bubble can lead to time scales for the evolution of $\psi$, that are of the same order of magnitude as the electron wavefunction, $\phi$. In this work, we will be predominantly concerned with an electron that remains in the ground state without any splitting of the bubble. Under such conditions, the adiabatic theorem can then be exploited to study the dynamics of the superfluid-bubble complex. In particular, for an electron that is initially in its ground state, we expect the electron to remain in its lowest energy level. This allows us to reformulate our original problem as
\begin{align}\label{Eq:GP_ad_1}
i \frac{\partial \psi}{\partial t} =-\frac{1}{2}\nabla^2 \psi+\frac{\gamma}{2}|\phi_g|^2\psi +\frac{1}{2}|\psi|^2\psi \, , 
\end{align}
where $\phi_g$ corresponds to the ground state that is determined by finding the minimum energy, $E$, for which 
\begin{align}\label{Eq:SCH_ad_1}
E_e \phi_g = \left[ -\frac{1}{2\delta}\nabla ^2 +\frac{1}{2\delta}\left(\zeta^2|\psi|^2+yQ\right) \right]\phi_g \, ,
\end{align}
and Eq.~\eqref{Eq:normphi} are satisfied. Since the contribution of the electron wave-function in Eq.~\eqref{Eq:GP_ad_1} is given by the squared modulus $|\phi_g|^2$, the evaluation of the dynamical and the geometrical phases turns out to be unimportant in studying the dynamics of an electron bubble in a superfluid within the adiabatic approximation.

%\subsection{Numerical results \label{Sec:Numerical results}}
%\subsubsection{Parameters for the simulations}
%All the numerical simulations that we will present 
The system of equations presented above in the adiabatic approximation were solved in a periodic domain using the algorithm described in Appendix \ref{App:Num}. This was implemented on a Tesla K40 NVIDIA graphics card. We modelled a flow in a channel of length $L_x=1024$, $L_y=128$ and $L_z=128$ with resolution set to $\Delta x = \Delta y = \Delta z= 1$. Given the localised nature of the electron wave-function, this was resolved on a smaller domain of dimensions $L_x=128$, $L_y=128$ and $L_z=128$, and constrained to lie within the central region of the channel as illustrated in Fig.~\ref{fig:chaos_large}. To initialise an electron in its ground state, at the beginning of each run an initial condition correctly describing the lowest energy state for the system of Eqs.\ \eqref{GP_ad} and \eqref{SCH_ad} with $Q=0$ is needed. We accomplish this by initializing the wavefunctions to correspond to the solution of an electron bubble trapped within a spherical cavity with hard walls. This initial guess is then relaxed by using the so-called gradient flow method \cite{Eloranta2007} which consists of integrating Eqs.~\eqref{GP_ad}-\eqref{SCH_ad} in imaginary time. More details are given in Appendix \ref{App:IC}. Upon recovering the desired initial condition, Eq.~\eqref{Eq:GP_ad_1} is then integrated in real time while Eq.~\eqref{Eq:SCH_ad_1} is solved in the presence of an applied electric field corresponding to $Q \ne 0$. The time step used for integrating the GP equation was $\Delta t=0.01$ while the step used for the gradient flow method to find the ground state of the time-independent Schr\"{o}dinger equation was set to $\Delta\eta=0.0001$. The gradient flow method was applied at each step until the $\mathsf{L}^2$ norm of $\phi_g$ appearing in Eq.~\eqref{Eq:SCH_ad_1} satisfied the threshold $\text{Err}<10^{-6}$. We have checked that such values were sufficient to accurately capture the coupling between the electron and the superfluid wavefunctions. Tests carried out using smaller values of the threshold did not affect our results significantly. Throughout the numerical solution procedure, we allowed the bubble to evolve over 100 time steps before shifting the entire fields, such that the bubble was re-centred within the channel using the procedure described in Appendix \ref{App:Num}. %\S~\ref{Sec:Dynamical evolution}

\section{Results \label{Sec:Results}}
%The first problem addressed in this section is the study of the motion of a free electron in helium-4 i.e. one that is not trapped on a vortex line.
The transport of negative ions in liquid helium has been the subject of experimental investigation for some time in order to understand the different forms of drag that can arise on an object moving through the superfluid \cite{Bruschi1966,Allum1976}. It is now well established that, at finite temperatures, the velocity of an ion is limited by the scattering of thermal excitations which consist of rotons and phonons. As the temperature is lowered below $0.7-0.8$K, the density of rotons falls off rapidly leaving phonons as the key remaining thermal excitations that interact ballistically with the ion. In the limit of $T=0$K, the density of phonons and rotons becomes so small that the kinetic energy transferred to the ion by the applied electric field can not be dissipated by interaction with thermal excitations alone. There is compelling experimental evidence which indicates that the ion can accelerate until it attains a critical velocity for the nucleation of vortex rings\cite{Rayfield1963,Bruschi1966,Allum1976,Zoll1976,vanDijk1977}. Depending on the strength of the applied electric field, it is believed that the ion can become either trapped on the core of a nucleated vortex ring, or it can continue to shed a stream of rings while undergoing intermittent vortex recapture events. This mechanism of nucleation of vortex rings is believed to provide a significant contribution to the drag experienced by the ion. 

Nancolas and McClintock \cite{Nancolas1982} showed that such a transition in which the ion is captured by the nucleated ring can be suppressed by operating at high pressures and by applying a sufficiently high electric field. In this regime, the ion can exceed Landau's critical velocity, which corresponds to the velocity at which rotons should be excited. They also demonstrated that as the operating pressure is lowered below 16 bar, the experimental data of Nancolas {\em et al.} showed a clear drop in the drift velocity of the ion (see Fig.\ 2 of Ref.\ [\onlinecite{Nancolas1985}]). They attributed this behaviour to the continuous generation of vortex rings in which the ion can undergo intermittent vortex capture events.

Given that many of the processes occur on scales that are impossible to observe directly, many of the assertions that have been made from existing experimental measurements have not been confirmed. Moreover, to date, no direct modelling has been performed to reinforce the conclusions drawn from data collected from measurements. In particular, as pointed out above, much of the modelling that has been carried out has been based on simplifying assumptions that are often unphysical. In addition, there has not been a systematic study of the response of the ion to different applied electric fields.
%shows that the emission of rotons is no longer the main mechanism affecting the velocity of the ion.} Hence a natural question that arises is what limits the velocity of the ion in this parameter range. Nancolas {\it et al.} 
%{\color{red}\cite{Nancolas1985} speculated about the emission of a stream of vortex rings, 
%{\color{red}remove comment on -  Guo and Jin \cite{Jin2010} proposed the emission of sound waves as new mechanism to dissipate kinetic energy.} 
\begin{figure}
 \centering
      \includegraphics[scale=0.4]{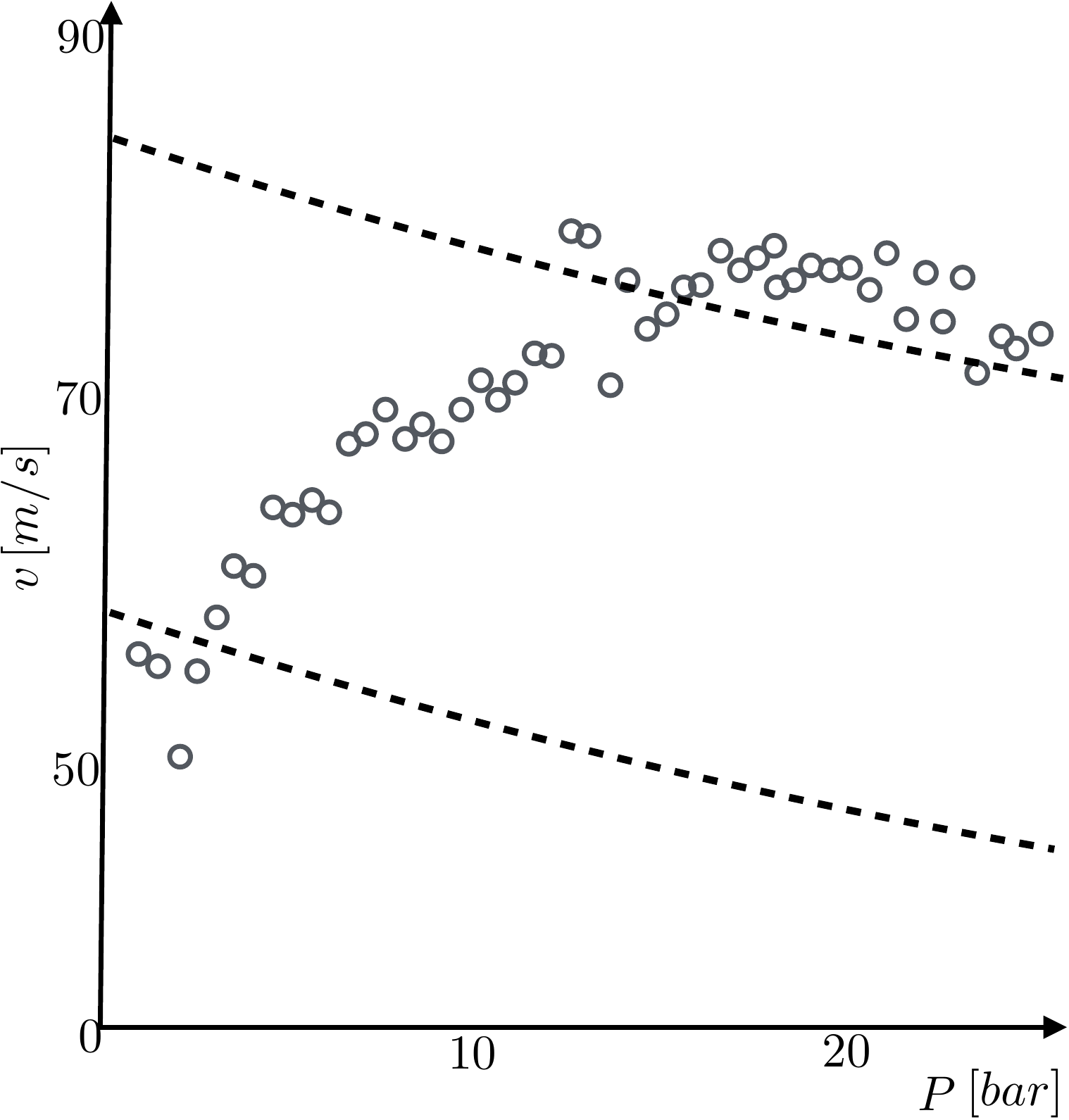}
 \caption{Plot of the drift velocity of negative ions  as a function of pressure in He II at 0.3 K moving under the influence of an applied electric field of $2.6 [MV/m]$. The lower dashed line represents the dependence of the Landau's critical velocity $v_L$ on the pressure while the upper line represents the expected drift velocity $v_D$ of the ion if the difference $v_D-v_L$ were to remain the same as the value measured at a pressure of 25 bar. (Data presented based on results published in Nancolas et al. \cite{Nancolas1985}) \label{fig:Nancolas}}
\end{figure}

\begin{figure*}
 \centering
      \includegraphics[scale=0.5]{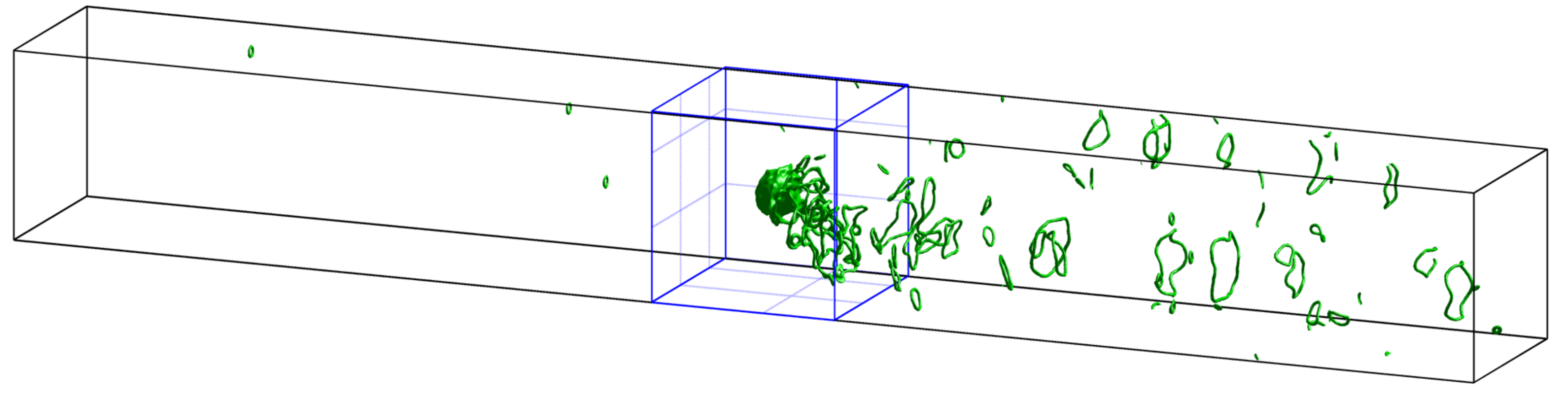}
      \caption{Isosurface plot corresponding to $|\psi^2|=0.3$ for an electric field $Q=5\times10^{-4}$. The figure shows the highly iregular generation of vortex rings that interact together to form a small vortex tangle behind the ion. The size of the computational domain used to numerically integrate the superfluid wave-function $\psi$ is shown in black while the extent of the domain used for the electron wave function $\phi$  is shown in blue.\label{fig:chaos_large}}
\end{figure*}

In order to resolve the questions concerning the nature of the dissipation mechanism at low temperatures and low pressures, we will numerically model the motion of an ion under different electric fields. Given that our model includes neither thermal excitations nor rotons, we will use the model to focus on how the nucleated vortex rings affect the motion of a bare electron bubble.
In the absence of any damping due to thermal excitations, the velocity of the ion is expected to increase linearly under the influence of an externally applied electric field. This should continue until a critical velocity $v_c$ is reached that coincides with the onset of nucleation of vortex rings. 
\begin{figure}
\hspace{-0.3cm}
 \centering
     \subfigure[Square]{
      \includegraphics[scale=0.25]{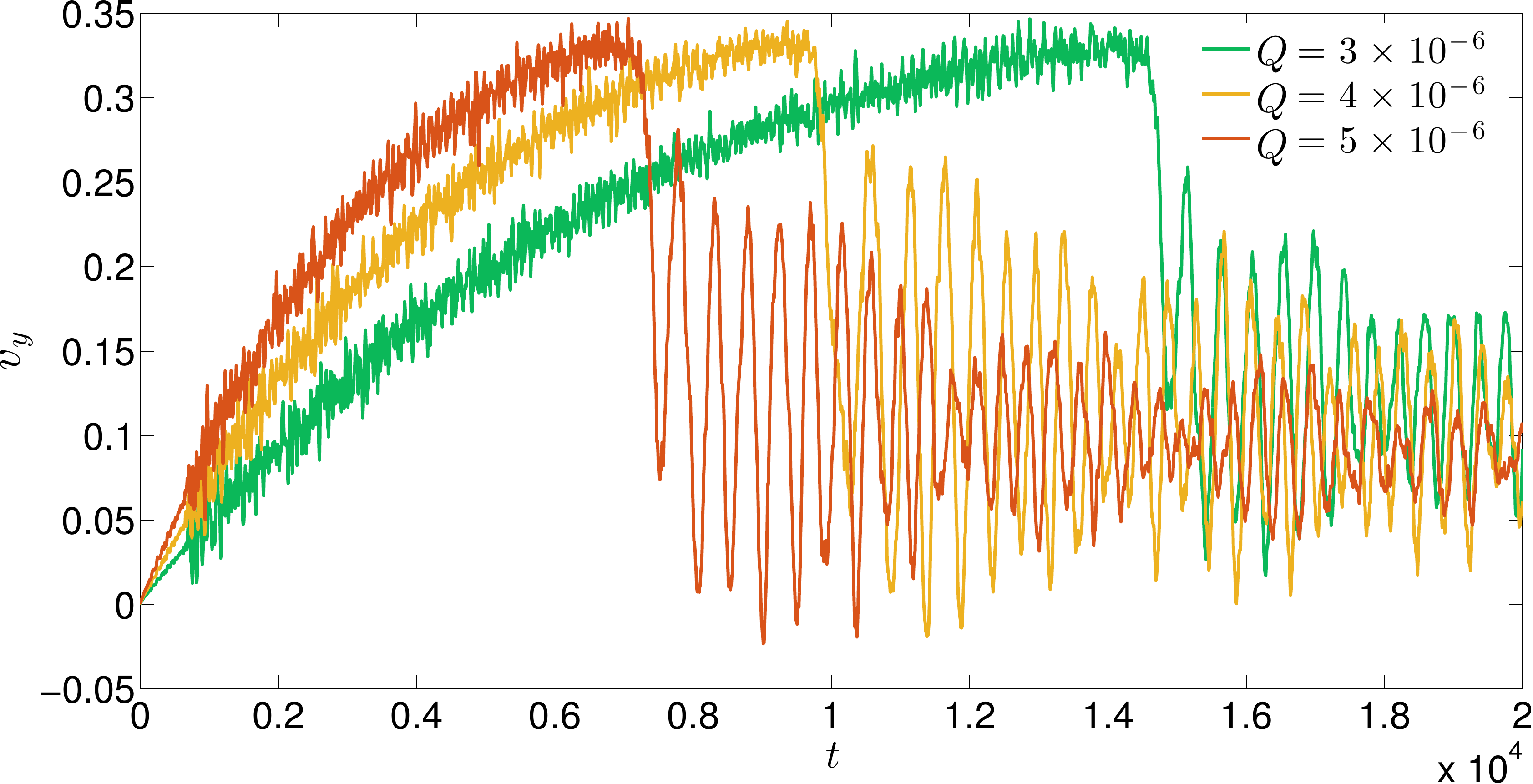}}
      \caption{Variation in time of the longitudinal velocity of the elecron bubble moving under the influence of electric fields of different strength. Each run shows an essentially discontinuous change in the velocity due to the capture of the ion by a nucleated vortex ring. After the sudden drop in velocity, the ion experiences large fluctuations due to Kelvin waves excited during the recapture process.\label{fig:small_E}}

\end{figure}
In Fig.~\ref{fig:small_E}, we present the time variation of the $y$-component of the velocity of the bubble that is estimated as
\begin{align}\label{Eq:vCM}
v_Y\simeq\frac{Y_{CM}(t+\Delta t_s)-Y_{CM}(t)}{\Delta t_s} \, ,
\end{align}
where $Y_{CM}$ represents the $y$-coordinate of the centre of mass of the bubble as defined in Eq.~\eqref{Eq:yCM}. Since the ion does not experience any drag during the early stages of the dynamics, the acceleration of the ion will initially be governed by the equation %(in dimensional units)
%\begin{align}
%\dot{v_y} \left( \frac{m_h  a}{\sigma^2} \right) = eqQ \, ,
%\end{align}
%{\color {green} or
\begin{align}
\dot{v_y} = \frac{eq\sigma^2}{m_h a} Q =\frac{1}{2\delta}\frac{m_4}{m_h} Q \, .
\end{align}
%with $m_h$ is the effective mass of the ion~\eqref{Eq:EffM}.
By performing a linear fit within the time interval $0<t<500$ for the case with $Q=3\times 10^{-3}$ we obtained the bubble acceleration $\dot{v_y}=1.9\times 10^{-5}$ which corresponds to an effective mass of $184 m_4$ and to an effective radius of $16 \text{\AA}$ that is consistent with our estimates quoted in \S~\ref{Sec:Nondim}. 
%Both the effective radius and the effective mass of the bubble are smaller than the values measured in\cite{Huang2017} since our model is considering a higher value of pressure which causes the bubble to shrink.\\

\begin{figure}
 \centering
  	\subfigure[\;Electric field $Q=3\times10^{-6}$]{
      \includegraphics[scale=0.25]{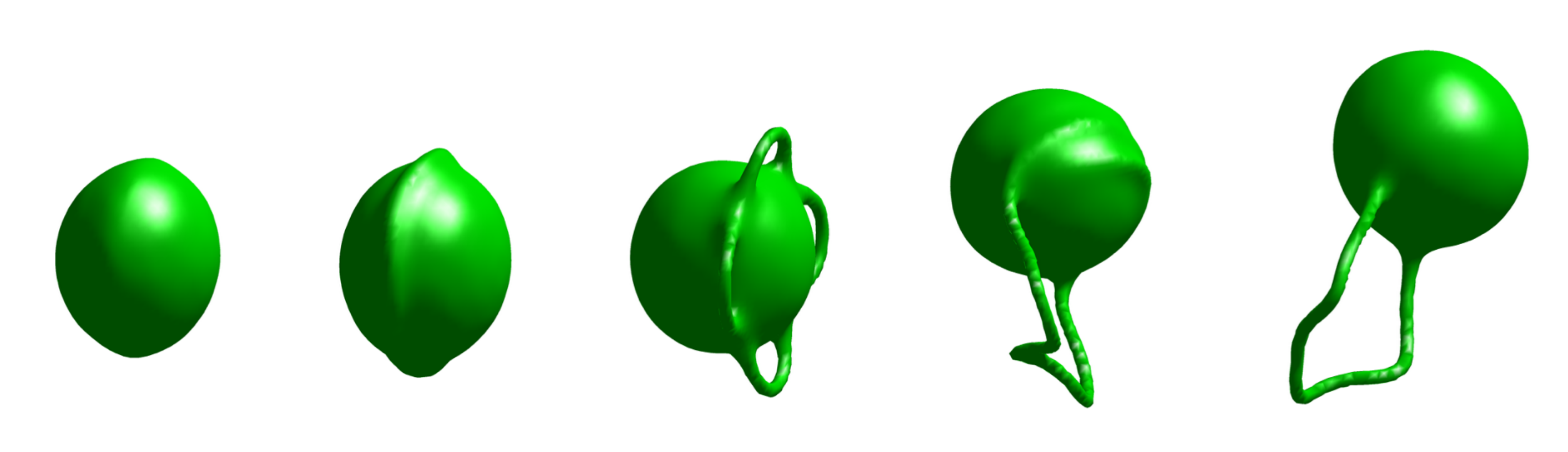}}
 \centering
   	\subfigure[\;Electric field $Q=10^{-5}$]{
      \includegraphics[scale=0.25]{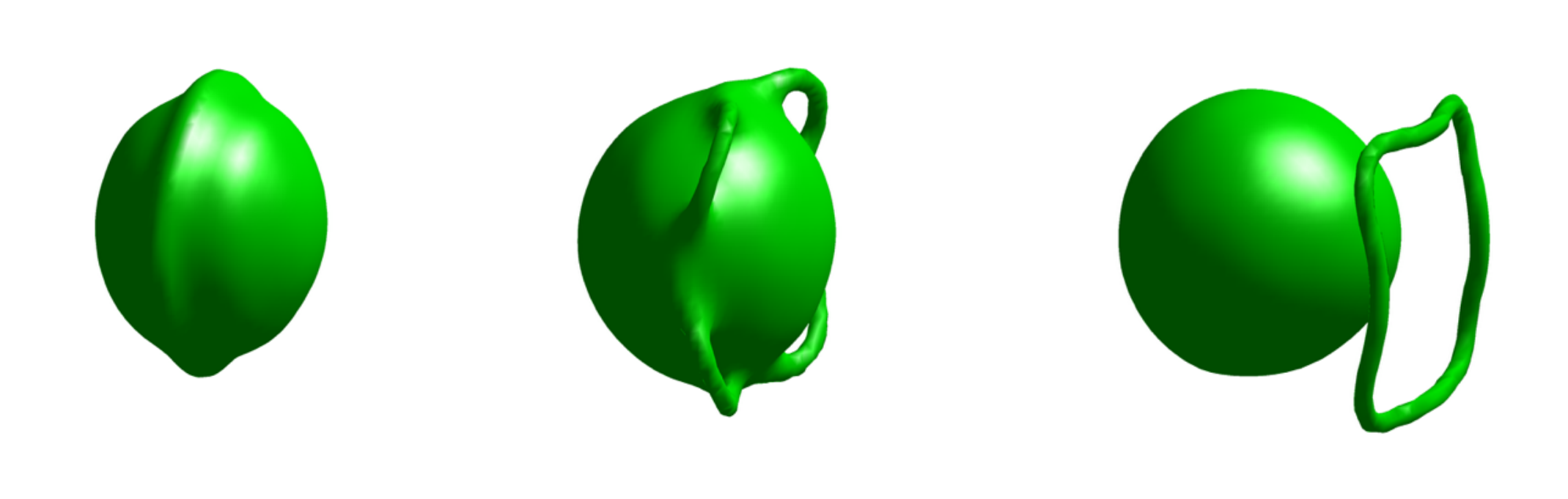}}
 \centering
   	\subfigure[\;Electric field $Q=2\times10^{-5}$]{
            \includegraphics[scale=0.25]{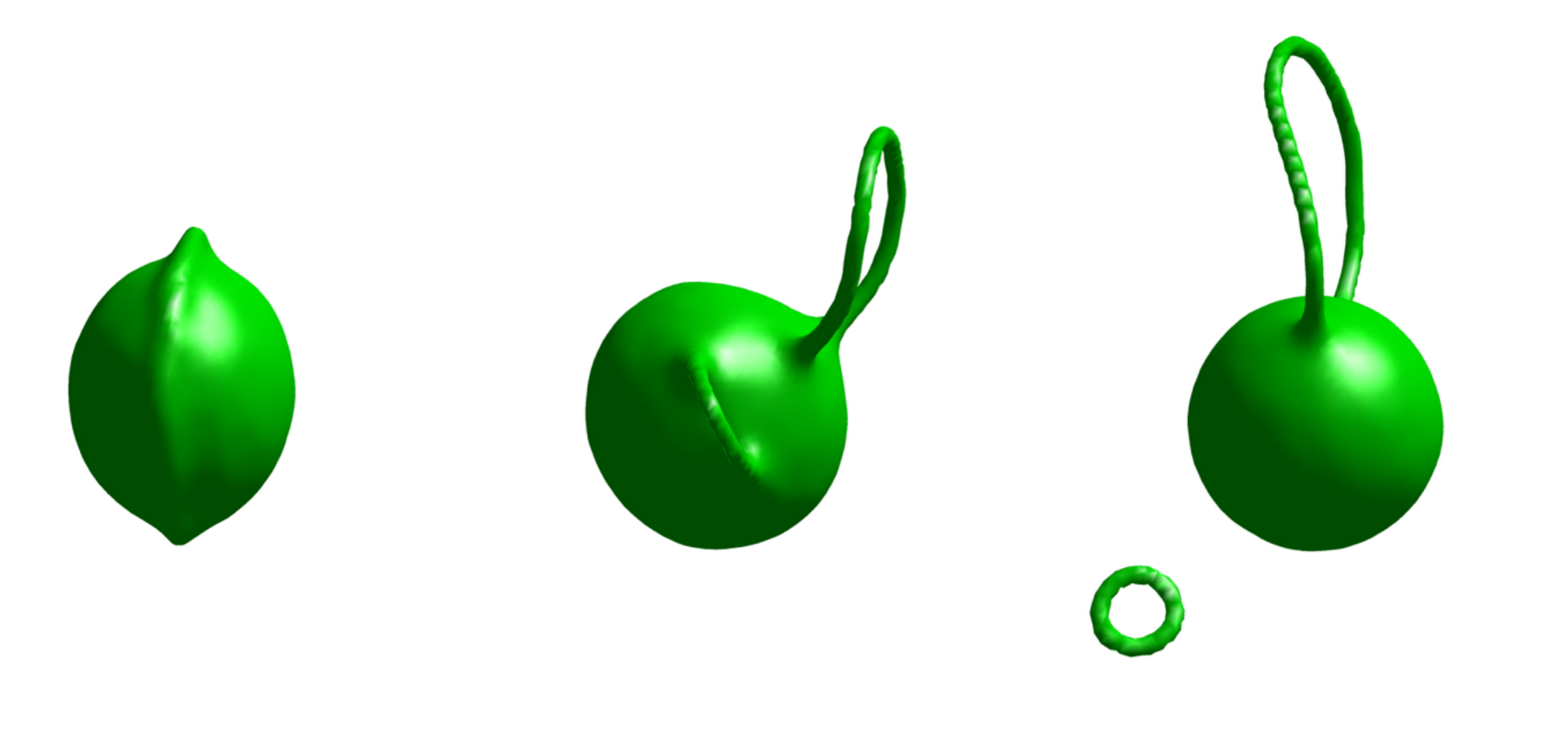}}
 \caption{Isosurface plot corresponding to $|\psi|^2=0.3$ of an electron bubble moving in the presence of constant applied electric fields. The sequence of images correspond to the times; (a) $t=12500$, $t=14750$, $t=15000$, $t=15500$ and $t=15750$; (b) $t=5250$, $t=5500$ and $t=5750$; (c) $t=16250$, $t=16500$ and $t=16750$ and they illustrate; (a) the transition from a free ion to a charged vortex ring; (b) the nucleation of a vortex ring according to the girdling model; (c) the transition to a charged vortex ring. \label{fig:capture_escape}}
\end{figure}

Following the initial linear growth, the velocity starts decreasing in time due to the deformations of the bubble. 
%{\color{red}which increases the effective mass of the bubble (*** need to be careful ***)}. 
When the velocity attains a critical value of $v_c\sim 0.32$, the bubble begins nucleating a vortex ring which subsequently reattaches to the ion. This process gives rise to the formation of a charged vortex ring\cite{Rayfield1964,Walmsley2014}. Details of this transition are illustrated in Fig.~\ref{fig:capture_escape}a. The transition to a charged vortex ring is associated with a sudden drop in the velocity of the bubble. During the recapture of the ion by the ring, sudden sideways motion of the ion occur that generate large perturbations on the ring. These fluctuations can be clearly seen in Fig.~\ref{fig:small_E} following the characteristic sudden drop in the velocity of the ion. We expect that the decay of these oscillations will be mediated by non-linear interactions of Kelvin waves that act to transfer energy to smaller scales until they are dissipated through emission of phonons\cite{Vinen2003}. At later times, the size of the charged vortex ring continues to increase with its velocity asymptoting to the self-induced velocity of a circular vortex ring \cite{Samuels1991,Winiecki2000}.

%self-reconnections for large amplitude perturbations\cite{Helm2011}, or

\begin{figure}
 \centering
      \includegraphics[scale=0.25]{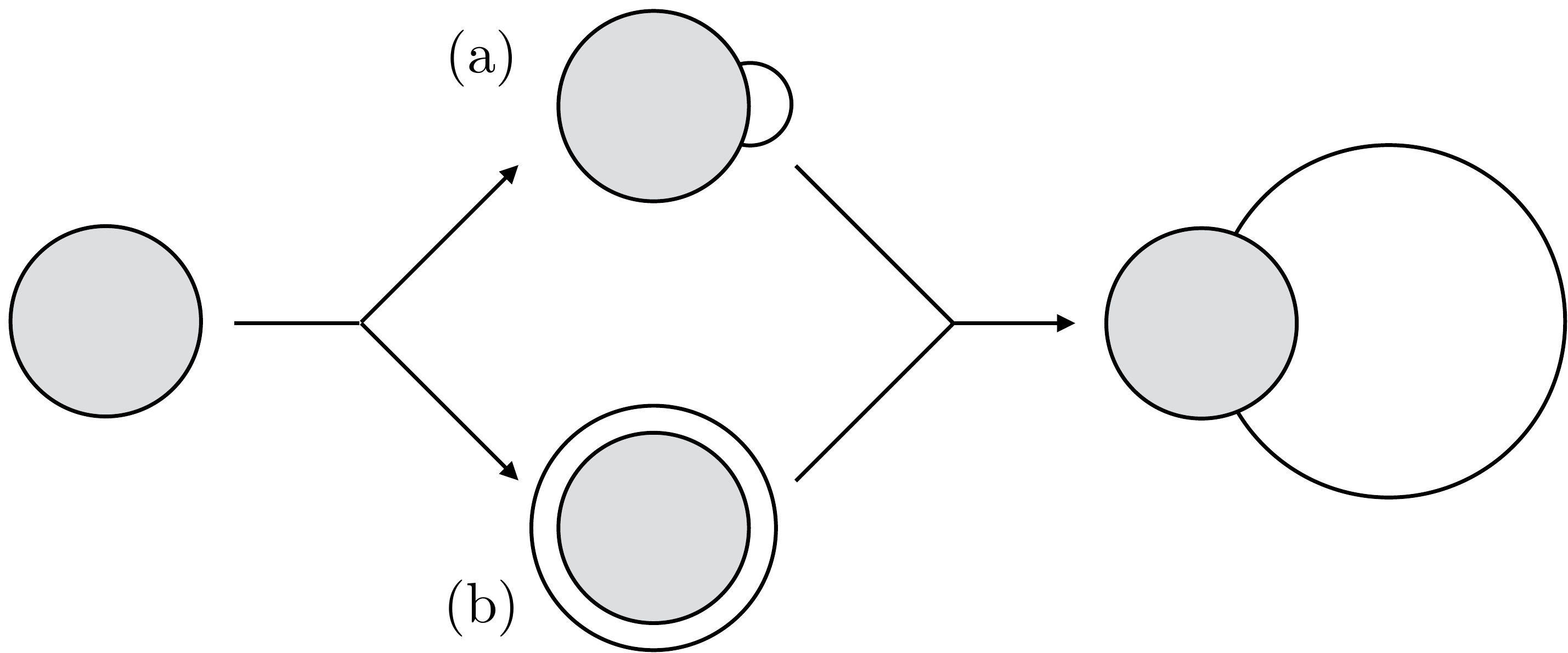}
 \caption{Schematic of the two alternative models describing the nucleation process: a)the peeling model; b) the girdling model. The arrows indicate the direction of time (image from Bowley et al. \cite{Bowley1982}).\label{fig:nuc_scheme}}
\end{figure}

%several theories have been proposed in the past to explain 
%{\color{red}(*** Need to include new references and use ideas discussed in book by Borghesani ***)}
The mechanism by which vortex nucleation occurs is a subject that has attracted much attention in the past and is one that has lead to several different explanations. In particular, we recall the works by Bowley {\it et al.} \cite{Bowley1982} and Murihead {\it et al.} \cite{Muirhead1984}. In these works, two different competing mechanisms were presented that came to be known as the girdling model and the peeling model, respectively. In the former case, a vortex ring detaches from the equator of the bubble, whilst in the latter case a vortex ring grows out from a small vortex loop that is attached asymmetrically to the bubble. A schematic plot of these two models is presented in Fig.~\ref{fig:nuc_scheme}. 
Our simulations corroborate observations made in previous numerical studies\cite{Berloff2000,Winiecki2000} and reveal that as an electron bubble is accelerated by a constant electric field, vortex nucleation is initiated by the emergence of a perfectly circular ring along the equator. This is also in agreement with theoretical models proposed by Schwarz and Jang\cite{Schwarz1973}, and by Bowley\cite{Bowley1984} for the initial stages of the process of vortex nucleation. However, it appears that this scenario which is consistent with the mechanism depicted in Fig.\ \ref{fig:nuc_scheme}b is inherently unstable to azimuthal perturbations. Consequently, as the ring begins to detach, it does not preserve the axisymmetry and leads to the formation of several smaller loops detaching from the bubble. 
Bernoulli effects associated with the nucleation of the vortex ring results in a pressure drop which causes the ion to become more susceptible to perturbations that causes the ion to begin to move in the transverse direction.
%Eventually, the bubble becomes susceptible to instabilities that give rise to transverse motion of the bubble and 
Consequently the ion moves off-centre with respect to the axis of the nucleated ring and is thus recaptured. 

%In contrast to previous numerical studies \cite{Berloff2000,Winiecki2000} we conclude that an electron bubble, accelerated by a constant electric field, nucleates a vortex ring according to the peeling model. In Fig.~\ref{fig:capture} it is possible to see how small loops, initially attached to the bubble, merge together forming a charged vortex ring with a large Kelvin waves on it.\\

We note that for these low electric fields, the nucleation always takes place at the critical velocity $v_c\sim0.32$ [\onlinecite{Berloff2001}]. Such a value of the critical velocity can be explained in terms of the motion of a sphere in an incompressible fluid as discussed by Berloff {\it et al.}\cite{Berloff2001} and Frisch {\it et al.} \cite{Frisch1992}. By working within a potential flow approximation of a classical fluid, it is known that the flow around such an object has a maximum velocity at the equator equal to $3/2U_{\infty}$, where $U_{\infty}$ is the velocity in the far-field. According to [\onlinecite{Berloff2001,*Frisch1992}], when $U_{max}$ matches the speed of sound $c=(1/\sqrt{2})$, that is set by the dispersion relation of the superfluid, vortex nucleation occurs. Small corrections due to the deformations of the bubble during its motion can modify the value of the critical velocity. This has been calculated in [\onlinecite{Berloff2001}] and it was found that 
$U_{max}=c$ when $v_c\sim0.34$, which turns out to be in good agreement with our observed numerical value.

\begin{figure}
 \centering
 \subfigure[\;Intermediate values of electric field]{
      \includegraphics[scale=0.36]{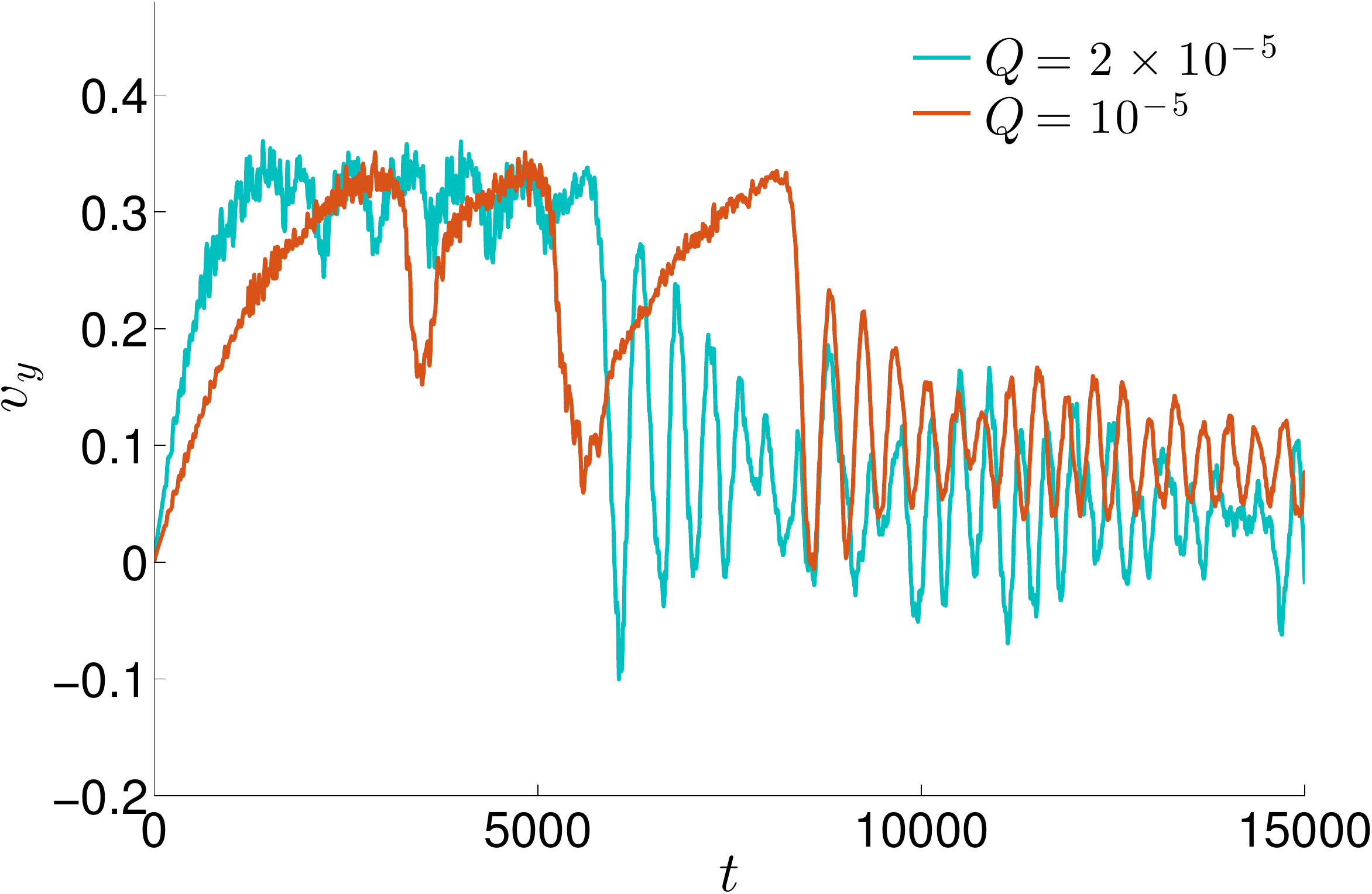}}
% \caption{Plot of the time evolution of the velocity of an electron bubble for two different values of the applied electric field. The transtion to a charged vortex ring take  place after the emission of several vortex rings. The time between two subsequent nucleation processes decreases with the increasing strength of the electric field.\label{fig:velocities_inter}}
 \centering
  \subfigure[\;High values of electric field]{
      \includegraphics[scale=0.36]{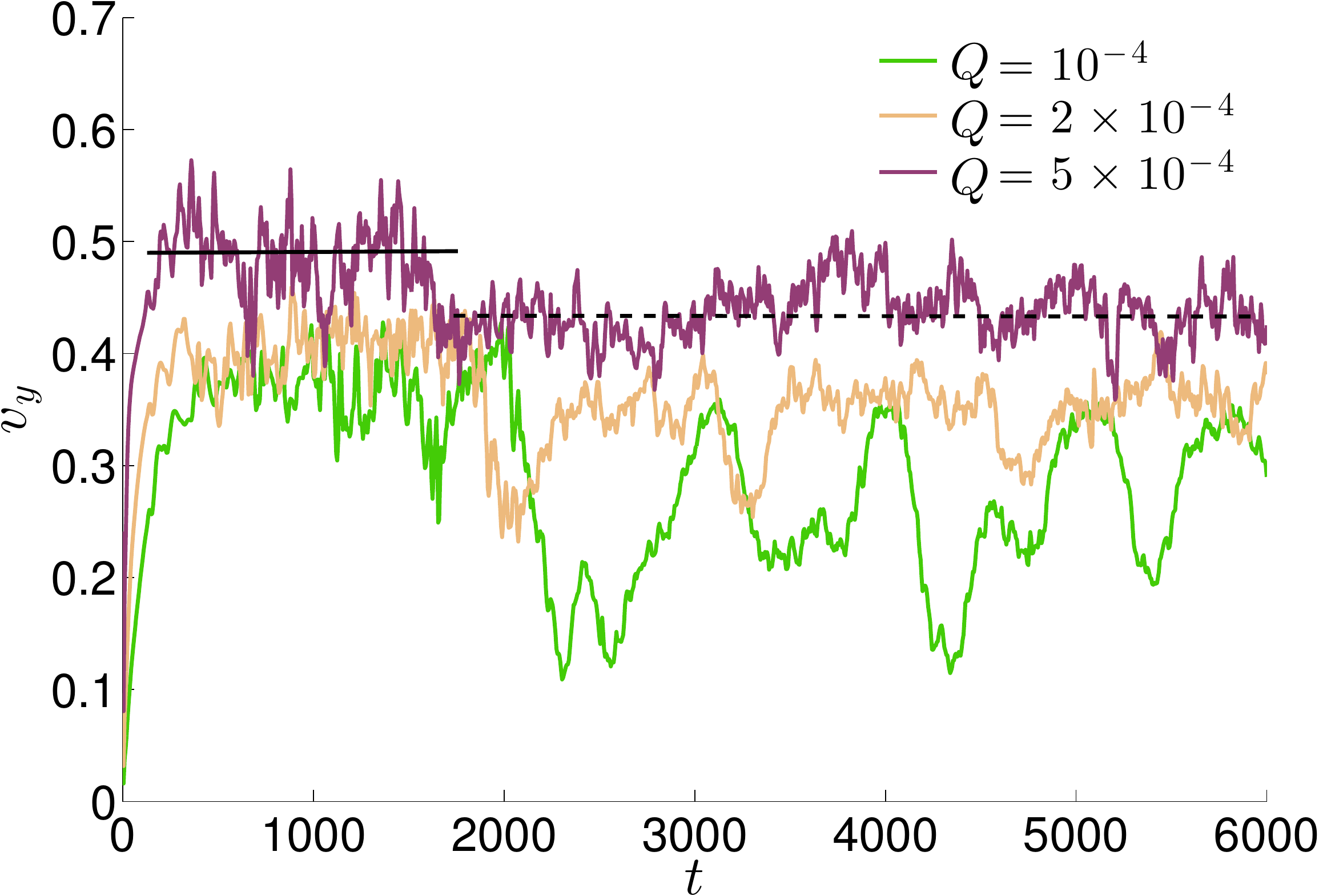}}
 \caption{Plot of the time evolution of the velocity of an electron bubble for different values of the applied electric field; (a) the transtion to a charged vortex ring takes place after the emission of several vortex rings. The time between two subsequent nucleation processes decreases with the increasing strength of the electric field; (b) for even higher electric fields, we observe that the velocity plateaus at different values depending on whether (solid line) or not (dash line) the axis-symmetry is broken.\label{fig:velocities_field}} 
\end{figure}

For higher electric fields, a markedly different behaviour is observed in that the ion enters a regime where a vortex ring is nucleated but manages to fully escape from the ion (see Fig.~\ref{fig:capture_escape}b). 
%\begin{figure}
% \centering
%      \includegraphics[scale=0.25]{figs/asim_capture-crop.pdf}
% \caption{Isosurface plot corresponding to $|\psi|^2=0.3$ of an electron bubble moving in the presence of a constant applied electric filed $Q=2\times10^{-5}$. The sequence of images correspond to times $t=16250$, $t=16500$ and $t=16750$ respectively, and show the transition to a charged vortex ring.\label{fig:asim_capture}} 
%\end{figure} 
The deflection of the trajectory of the ion leads to the development of chaotic dynamics. For example, for even higher electric fields we observe in Fig.~\ref{fig:capture_escape}c that transverse motion of the bubble can lead to the formation of two vortex loops with different sizes, the smaller of which detaches from the ion while the larger one captures the bubble.  The detachment of the ring from the ion leads to an intermittent signal in the magnitude of the longitudinal velocity of the ion as illustrated in Fig.~\ref{fig:velocities_field}a.
This clearly indicates the nucleation of several vortex rings that is evident from the abrupt fall off in the velocity of the ion that takes place at different instants in time. In particular, for $Q=10^{-5}$ at $t=3000$ the ion reaches the critical velocity $v_c$, nucleates a vortex ring with a consequent drop-off in the velocity. Thereafter, the ion accelerates until it again reaches the critical velocity, $v_c$, and the system cycles again through the same sequence of events. Eventually, after the nucleation of several vortices, the ion finally becomes trapped, resulting in a charged vortex ring with the velocity fluctuating around the value $v_y\sim0.1$. As can be seen from Fig.~\ref{fig:velocities_field}a, the time between two subsequent vortex nucleation processes decreases with the increasing strength of the electric field.\\

We recall that Nancolas {\it et al.}\cite{Nancolas1982}, suggested that the transition to a charged vortex ring can be suppressed by applying sufficiently high electric fields. 
\begin{figure}
 \centering
      \includegraphics[scale=0.25]{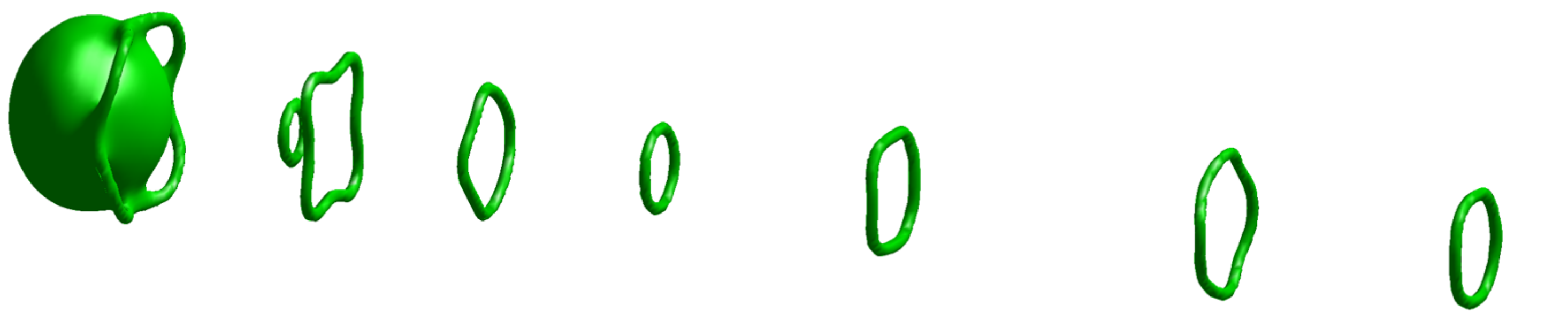}
 \caption{Isosurface plot corresponding to $|\psi^2|=0.3$ for an applied electric field $Q=10^{-4}$. The figure shows the emission of a stream of vortex rings. The smaller rings are nucleated at earlier times and they travel faster towards the ion by undergoing a leapfrogging motion with the array of rings nucleated at later times. \label{fig:leapfrogging}}
\end{figure}
In Fig.\ \ref{fig:leapfrogging} we show a stream of vortex rings having more or less the same size as the bubble for $Q=10^{-4}$. We have found that these nucleated rings interact together giving rise to a leapfrogging type behaviour (see Supplemental Material for explanatory movies). In particular, vortex rings nucleated at earlier time can be slowed down and eventually propelled toward the ion until they scatter off the ion. Because of the collective motion of the nucleated rings, the velocity of the ion initially exceeds the critical value $v_c$ but subsequently enters a regime characterized by highly chaotic dynamics with irregular vortex shedding (see Fig~\ref{fig:chaos_large}).
Upon increasing the strength of the electric filed, the frequency for the emission of vortex rings increases. As shown in Fig~\ref{fig:chaos_large}, for $Q=5\times10^{-4}$, the nucleation becomes so rapid that a small vortex tangle develops in the wake of the ion.\\
%\begin{figure}
% \centering
%      \includegraphics[scale=0.3]{figs/velocities_large_E1-crop.pdf}
% \caption{Plot of the time evolution of velocity of an electron bubble under different values of electric fields. The velocity is plateauing at different values depending on whether the axis-symmetry is broken (solid line) or not (dash line).\label{fig:large}}
%\end{figure} 

In Fig.~\ref{fig:velocities_field}b, we present the evolution of the velocity of the bubble for higher values of the electric field. The figure shows that the ion experiences two different regimes during its dynamics. More specifically, by analysing the case corresponding to $Q=5\times10^{-4}$ (purple line), it is possible to see that the velocity of the bubble initially plateaus at $v_y\sim0.5$ (see solid line). This value is associated with the axis-symmetric nucleation of vortex rings. Once the symmetry breaks down, the motion of the ion becomes chaotic and the value of the velocity significantly changes (see dash line). 
\begin{figure}
 \centering
      \includegraphics[scale=0.36]{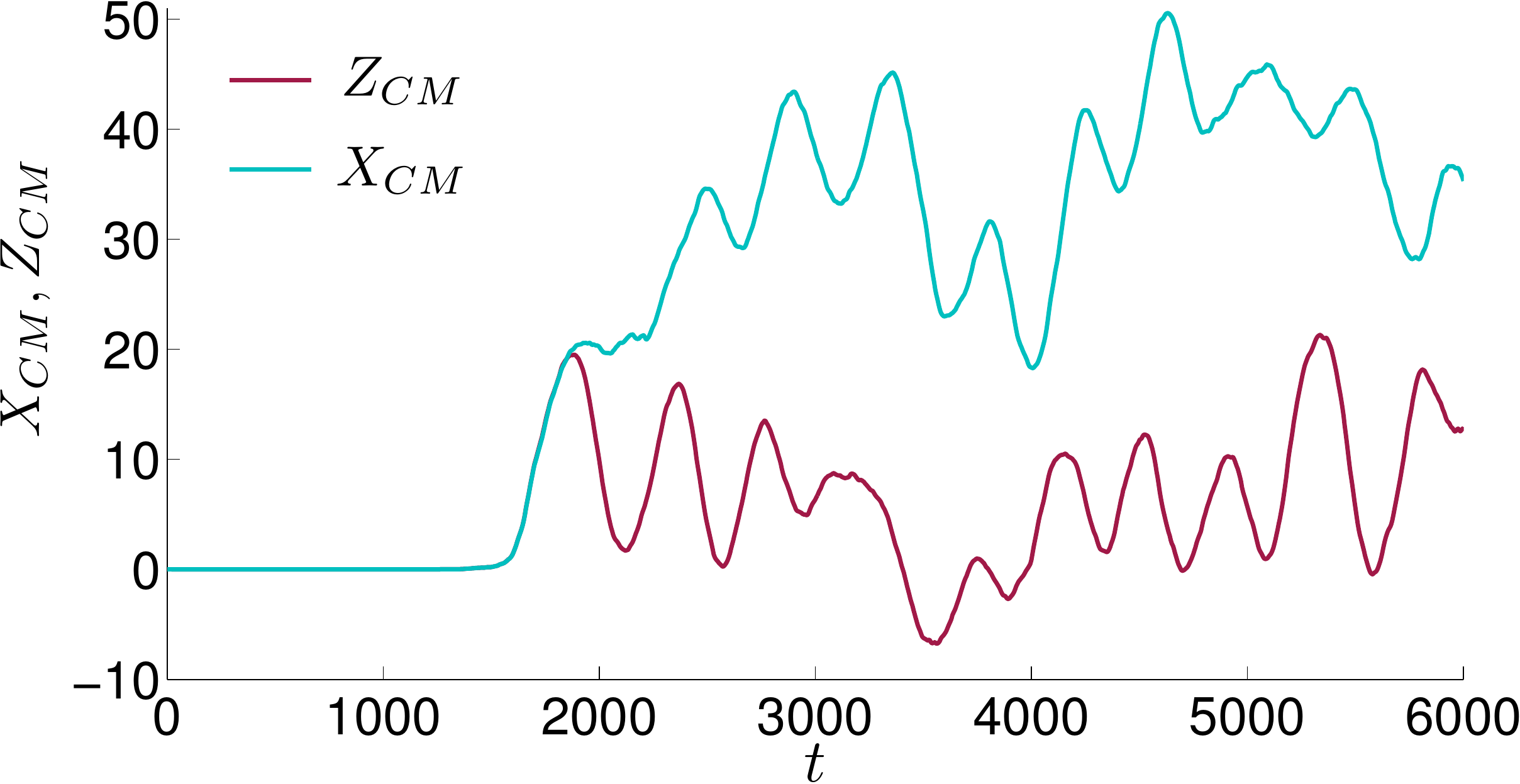}
 \caption{ Plot of the $x$ and $z$ coordinates of the centre of mass of the bubble. The fluctuating behaviour is the signature of the transition into a chaotic regime.\label{fig:x_z}}
\end{figure}
To detect the moment when the transition into a chaotic regime takes place, in Fig.~\ref{fig:x_z} we plot the $x$ and the $z$-coordinates of the centre of mass of the bubble denoted by $X_{CM}$ and $Z_{CM}$, respectively which are evaluated according to~\eqref{Eq:yCM}. The figure clearly shows that the onset of chaotic motion of the bubble occurs at $t_c\sim 1600$ which coincides with the transition from $v_y \sim 0.5$ to $v_y \sim 0.4$ seen in Fig.\ \ref{fig:velocities_field}b. A drift velocity for the bubble can be evaluated by averaging over time the velocity, $v_y$, of the bubble after the transition $t_c\sim 1600$ has occurred.
\begin{figure}
 \centering
      \includegraphics[scale=0.35]{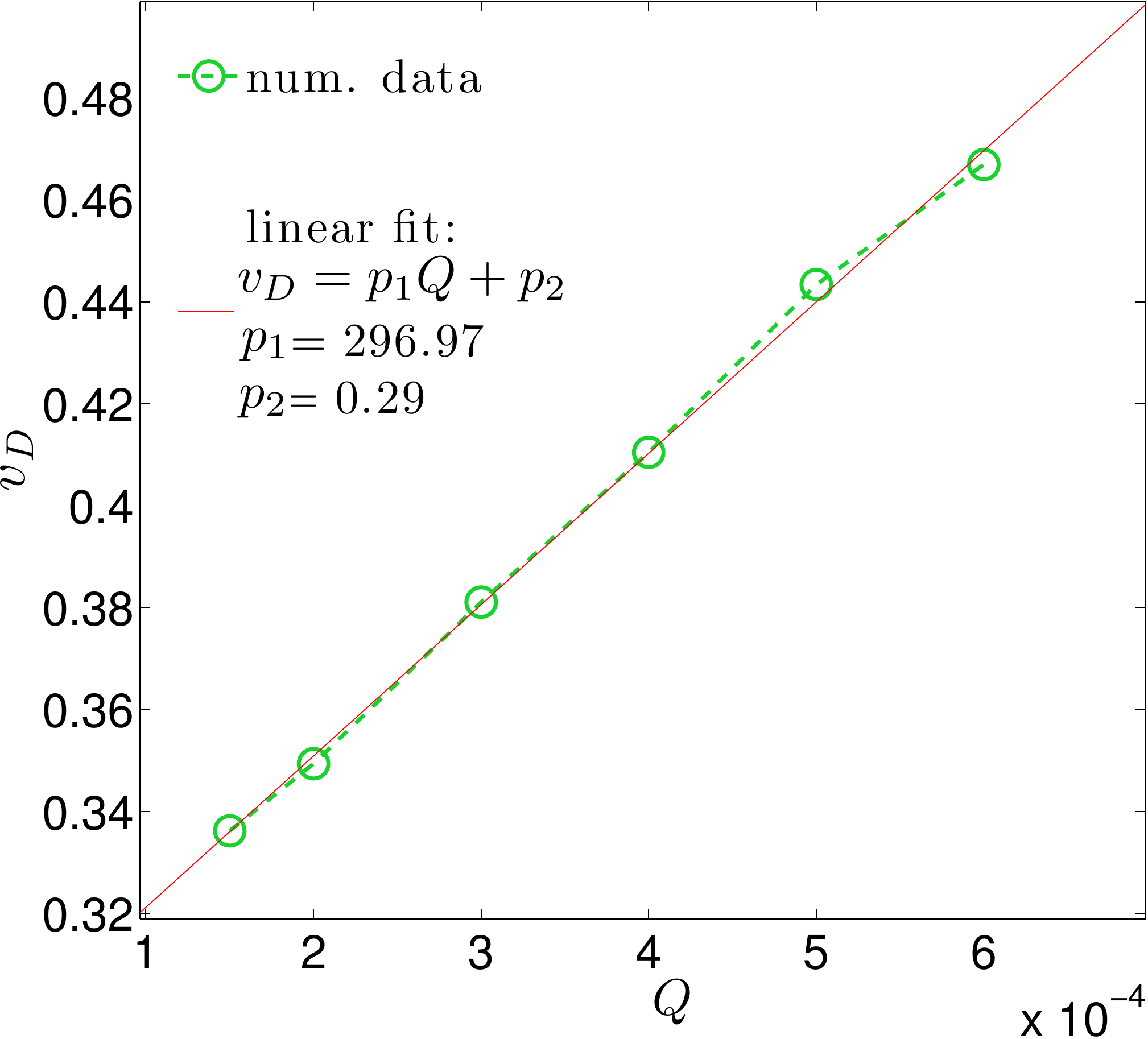}
 \caption{Plot of the drift velocity of an electron bubble under different applied electric fields. A linear relation between $v_D$ and $Q$ is found by fitting the numerical data with a first order polynomial $v_D=p_1 Q+p_2$ \label{fig:mobility}}
\end{figure}
In Fig~\ref{fig:mobility} we plot the drift velocities for different values of the applied electric field. 
An interesting observation that we make is the linear relationship that exists between $v_D$ and $Q$ within the range of values shown.

%{\color {blue}After evaluating the drift velocities for the bubble we can also estimate the quantum Reynolds number using the definition 
%\begin{equation}
%Re_s=\frac{2m_4(v_y-v_c)b\xi^2}{h\sigma}
%\end{equation}
%given by Reeves {\it et al.} \cite{Reeves2015}. Drawing analogies with the results obtained in [\onlinecite{Reeves2015}], where the transition to turbulence occurred for $Re_s\approx0.7$, we notice that such a value of Reynolds number corresponds to velocities of bubble larger than $v_y\approx 0.45$. This is indeed the case, as shown in Fig.~\ref{fig:chaos_large}, when $Q=5\times 10^{-4}$. For lower electric field the emission of vortex rings results in a more regular patter similar to what shown in Fig.~\ref{fig:leapfrogging}.}

We recall that the mobility of the ion is defined as $\mu_{\text phonon} = v_D/Q$ in the limit as $Q \rightarrow 0$. In experiments, this mobility is typically determined by the phonon limited drift velocity since a finite fraction of phonons is typically present in experiments at low temperatures that scatter off the ion and lead to a drag force. However, experiments also reveal that for higher electric fields, exceeding the critical velocity coinciding with the formation of charged vortex rings where the drift velocity of the ion is seen to rapidly fall off, another regime is encountered where the ion's velocity is seen to again increase with increasing field strength. This regime which is the one that is relevant to our numerical studies can be used to define a vortex nucleation limited mobility\cite{Phillips1974} given by $\mu_{\text ring} = v_D/(Q-Q_{\text cr})$. Here, $Q_{\text cr}$ coincides with the critical threshold of the electric field for which the drift velocity of the ion is seen to rise again. Using our results presented in Fig.\ \ref{fig:mobility}, we find $\mu_{\text ring} = 1.26\, \text{m}^2\text{s}^{-1}\text{M}^{-1}\text{V}^{-1}$.\\

Our value for the vortex nucleation limited mobility of ions, within the range of electric fields explored, can be compared against previously performed measurements of the same quantity. We note that previous work has studied the mobility of ions at high electric fields as a function of both pressure and temperature \cite{Phillips1974,Ellis1985}. We could not find data obtained for pressures that correspond directly to the conditions associated with the parameters used in our model. We will, therefore, consider two sets of data. The first is taken from [\onlinecite{Phillips1974}] which contains measurements for a pressure of $p=1$ MPa and taken over a range of temperatures that is of most relevance to our work. %The second is taken from [\onlinecite{Ellis1980}] which is date for $p=2.5$ MPa and $T=0.34$K. 
When comparing with experimental data collected at high pressures, consideration must be given to the fact that the mechanism that determines the maximum drift velocity of the ion is dependent on the pressure. In particular, as demonstrated in [\onlinecite{Ellis1980,Ellis1985}], roton pair creation is believed to be the dominant mechanism  above 10 bar, whereas vortex ring nucleation is the main mechanism below 10 bar. This is consistent with the observation that the Landau critical velocity for roton creation and the critical velocity for vortex ring nucleation both vary with pressure but the two velocities coincide at $p=10$ bar (see Fig.\ 1 in [\onlinecite{Ellis1985}]). The measurements presented in [\onlinecite{Phillips1974}] for $p=10$ bar are, therefore, most relevant for our simulations. Taking the measured vortex limited mobility presented in Fig.\ 19 of [\onlinecite{Phillips1974}], we find $\mu_E = 1.1\, \text{m}^2\text{s}^{-1}\text{M}^{-1}\text{V}^{-1}$ which is in remarkably good agreement with our value quoted above. 

To establish the sensitivity of these results with changes in the operating pressure and, more specifically, to quantify to what extent the emission of roton pairs affects the measured mobility, we have also analysed a second set of data presented in  [\onlinecite{Ellis1980}] for $p=2.5$ MPa and $T=0.34$K. In fact, in that work, the measured drift velocity $v_D(Q)$ of an ion had a discrepancy from the expected behaviour that is predicted if pair-roton emission is taken to be the main source of drag. As suggested in [\onlinecite{Nancolas1985}], such a discrepancy could be accounted for if one takes into account corrections arising from the emission of vortex rings. We have, re-analysed the experimental data to determine the measured mobility of the ion at this higher pressure. As can be seen from our the data included in Appendix \ref{App:Expt}, a linear relation can be identified between the measured drift velocity of the ion and the applied electric field. This allows us to obtain an experimental value of the measured mobility of $\mu_{E}=2.76 \,\text{m}^2\text{s}^{-1}\text{M}^{-1}\text{V}^{-1}$. This reveals that increasing the pressure increases the measured mobility. We note that at higher pressures, the radius of the electron bubble is reduced. Therefore, if vortices are nucleated together with pair-roton emission, the rings are expected to be significantly smaller in comparison to those formed at lower pressures. Despite these different physical effects, the measured mobility only increases by around a factor of 2. Therefore, given the simplicity of the model we have used, we are able to replicate within good quantitative agreement, the measured mobilities of the negative ions at high electric fields.

\begin{figure}
 \centering
      \includegraphics[scale=0.29]{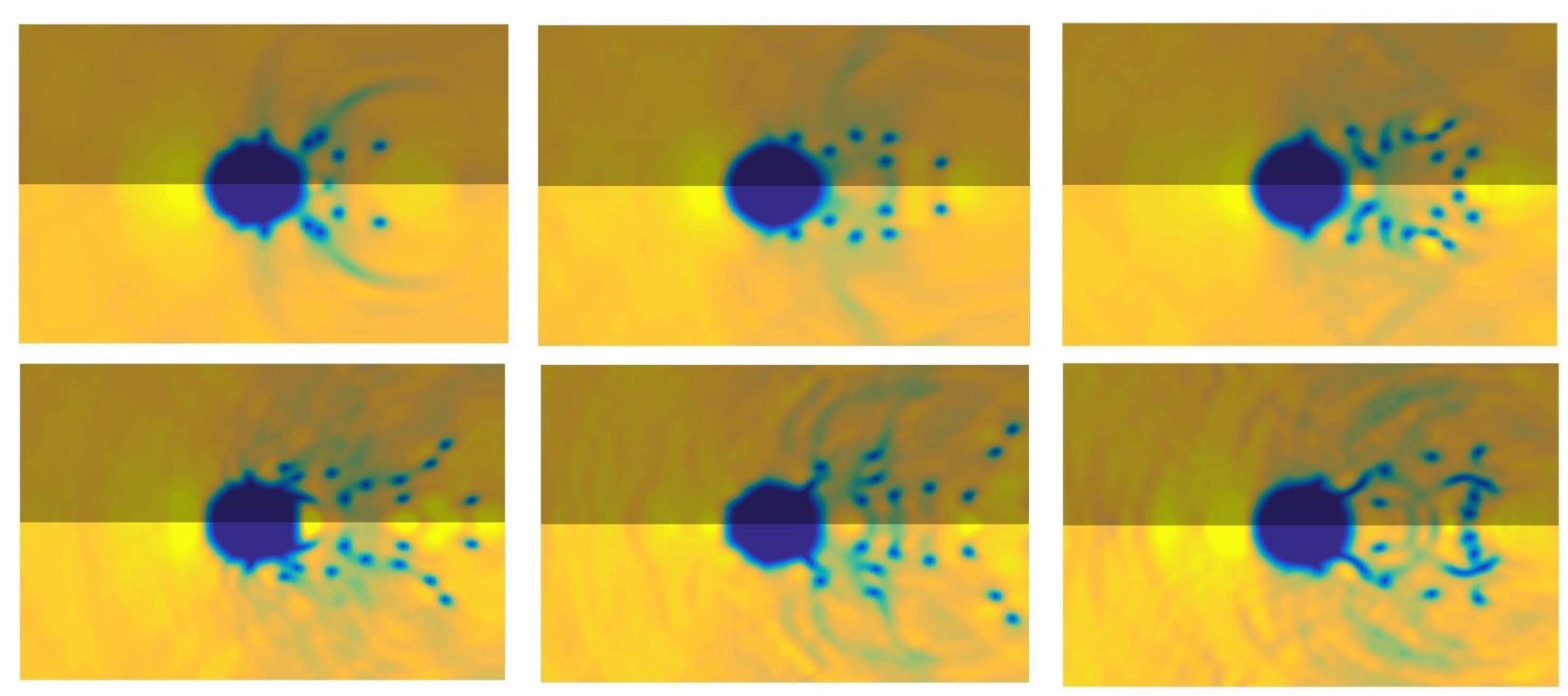}
 \caption{Plot of two slices corresponding to the plane $z=0$ and $x=0$ extracted from the 3D numerical domain. Dark areas represent depletions in the field $|\psi|^2$ while bright areas stands for high values of $|\psi|^2$. \label{fig:slices}}
\end{figure}

We end by noting that Guo and Jin \cite{Jin2010} have shown, using a density functional theory that emission of sound waves by disturbances of the bubble can provide a significant channel for dissipating energy. While we also observe the emission of sound waves as illustrated in Fig.~\ref{fig:slices}, the model used in [\onlinecite{Jin2010}] allows the correct equation of state for $^4$He to be used thereby providing a more accurate description of this dissipation mechanism. However, as shown in this work, their assumption of axisymmetry inhibits the transverse chaotic motion of the bubble that appears to be the dominant factor in determining the subsequent velocity of the ion at late times. Future work will aim to extend the 3D simulations we have performed to more realistic models such as the ones considered in [\onlinecite{Jin2010}]. This would permit a more quantitative determination of the different contributions to the drag force exerted on the ion.

%\begin{align}
%r_{CM}=\sqrt{x_{CM}^2+z_{CM}^2}
%\end{align}
%from the original straight trajectory of the ion $(0,y_{CM},0)$ in time.
%\begin{figure}
% \centering
%      \includegraphics[scale=0.3]{Bubble/Figs/velocities_check-crop.pdf}
% \caption{{\color {red} provvisory plot:}Different runs for the same value of electric field $E=10^{-5}$. The runs differ in the parameter of the simulation such as the length of the box, the size of the electron subgrid and the threshold value for imaginary time method. \label{fig:check}}
%\end{figure}
 %Experimental data shown in figure \ref{fig:Allum} confirm that at high $E$ the ion enter a regime where the drift velocity result mainly dominated by the emission of rotons. In our model we not able to reproduce such dynamics since the dispersion relation in the Gross-Clark model does not present any roton minimum. However we can investigate a different type of drag experienced by the ion due the generation of many vortex rings. 
%\begin{figure}
% \centering
%      \includegraphics[scale=0.45]{Bubble/Figs/leapfrogging-crop.pdf}
% \caption{Isosurfaces plots corresponding to $|\psi|^2=0.3$ of an ion pulled by a constant electric field $E=5\times10^{-5}$. From top to bottom we show a leapfrogging dynamics at time $t=2250$, $t=2300$, $t=2350$, $t=2400$ and $t=2500$
%\label{fig:leapfrogging}}
%\end{figure}
%In figure \ref{fig:leapfrogging} we show how 

\acknowledgements{The authors would like to thank Dr.\ P.~Walmsley, Dr.\ D.~Proment, and Dr.\ G.~Krstulovic for valuable discussions. HS acknowledges support for a Research Fellowship from the Leverhulme Trust under Grant R201540. Computations were carried out on the High Performance Computing Cluster supported by the Research and Specialist Computing Support service at the University of East Anglia.}

\appendix
\section{Spherical Cavity Model of Electron Bubble \label{App:Cavity}}
In order to gain further insight into the properties of the electron in the self-trapped bubble state, we will use the equations presented in \S~\ref{Sec:GrossClarkModel} to derive a simple model of a perfectly spherical cavity at equilibrium. This will help in identifying the key length scales that will arise in our problem and which need to be well resolved in our numerical simulations.

We begin by assuming that, the electron is in its $s$-state and is trapped within a perfectly spherical cavity of radius $b$ that represents the bubble state. For simplicity, the cavity is assumed to have an infinite depth. The lowest eigenvalue of Eq.~\eqref{Eq:H_SCH} will then be given by
\begin{align}\label{Eq:KinEn_Sphere}
E_q=\frac{\hbar^2 \pi^2}{2 m_e b^2} \, .
\end{align}
This expression corresponds to the quantum mechanical energy associated with the zero-point motion of the electron. Another key contribution to the total energy of the electron bubble and superfluid system is one that arises from the (non-linear) interaction part of the GP Hamiltonian~\eqref{Eq:GPHAM}. From this term, we can determine the work required to carve out a cavity within the superfluid due to the pressure field $p=V_0 \rho^2/2m_4^2$ for a spherical cavity. This contribution to the energy is given by
\begin{align}
E_V=pV=\frac{4\pi b^3}{3}p=\frac{2\pi b^3V_0\rho^2}{3 m_4^2} \, .
\label{Eq:E_V}
\end{align} 
The third principal contribution to the total energy of the system that is associated with the electron bubble state is given by the kinetic energy term in Eq.~\eqref{Eq:GPHAM}
\begin{align}\label{Eq:T}
\frac{\hbar^2}{2m_4}\int|\nabla \psi|^2 \d^3\vv{x}\sim \frac{2\pi \hbar^2\rho}{3m_4^2a^2}[(a+b)^3-b^3] \, .
\end{align} 
Here, $a$ is the healing length that sets the length scale over which the density of the fluid rapidly falls off from its far-field value (see [\onlinecite{Gross1966}]). For experimentally relevant parameters, we can assume that $a\ll b$. The dominant contribution to Eq.\ \eqref{Eq:T} is then proportional to the area of the bubble and we can interpret this term as the energy associated with the surface tension, $T$, of the cavity wall that can be expressed as
\begin{align}
E_T = 2\pi \frac{\hbar^2 b^2 \rho}{m_4^2 a} \equiv 4\pi T b^2 \, . 
\end{align}
We note that in [\onlinecite{Gross1966}], a slightly more accurate estimate of the surface tension was obtained by using a tanh profile to describe the superfluid profile at the boundary of the bubble.
In the cavity model of the electron bubble, the wavefunctions for the electron and the superfluid do not overlap and hence the interaction term given by Eq.\ \eqref{Eq:H_GP_e} does not contribute. The total energy for the electron bubble-superfluid system is then given by
\begin{align}\label{eq:Energy_Bubble}
E=E_q+E_V+E_T=\frac{\hbar^2\pi^2}{2 m_e b^2}+\frac{4\pi b^3}{3}p+4\pi Tb^2.
\end{align}

%\begin{equation}\label{eq:Energy_Bubble}
%E=\frac{\hbar^2\pi^2}{2 m_e b^2}+\frac{2\pi V_0 \rho^2 b^3}{3M^2}+4\pi Tb^2,
%\end{align}
%Here the first term arises from quantum mechanical zero-energy motion of a particle of mass $m_e$ in an infinite spherical potential well of radius $b$.
%The second term is the work needed to dig a spherical cavity of radius $b$ in a superfluid under the pressure $P$.
%%with $V_0$ being the two body short-range repulsive potential between helium particles of mass $M$. 
%The last term represents the interface energy between the electron and the superfluid with $T$ being the surface tension. 
%A more extensive discussion of this model was provided by Gross \cite{Gross1966} where the fermion-boson interaction in~\eqref{Eq:GC_Htot} was modelled using a hard-core interaction potential.}

Using this model, we can now estimate the radius $b$ of the electron bubble and subsequently its hydrodynamic mass $m_{h}$ \cite{Batchelor1967}. Since the electron mass $m_e$ is much smaller than the mass of the $^4$He atom, $m_4$, with $\delta={m_e}/{m_4}\sim1.4\times 10^{-4}$,
%\begin{equation}
%\delta=\frac{m_e}{m_4}\sim1.4\times 10^{-4},
%\end{align}
the effective mass of the bubble $(m_e+m_h)$ can then be approximated by its hydrodynamic mass which is given by
\begin{align}\label{Eq:EffM}
m_h=\frac{2}{3}\pi \rho b^3.
\end{align}
At zero pressure, Eq.~\eqref{eq:Energy_Bubble} can be used to evaluate the radius of the bubble that minimizes the electron energy $E$; this gives
\begin{align}
b=\left(\frac{\pi\hbar^2}{8m_e T} \right)^{1/4} \, .
\end{align}
Using typical measured values of parameters for liquid helium at zero temperature, such as the surface tension of bulk helium\cite{Roche1997}, $T=375 \, \mu \text{J m}^{−2}$, and the liquid density $\rho= 0.145\, \text{g/cm}^3$, we can finally estimate that the effective radius is $b=18.91 \text{\AA}$ whereas the mass $m_h=309\,m_4$ for an electron bubble at zero pressure.\\

For non-zero pressure, it is possible to estimate the radius of the bubble by using the method of dominant balance under the condition that $\delta\rightarrow 0$. We begin by noting that a stationary value for the energy~\eqref{eq:Energy_Bubble} is given by the solution of 
\begin{align}\label{eq:Min_Energy_Bubble}
\frac{d E}{d b}=
-\frac{\hbar^2\pi^2}{m_e b^3}+4 \pi b^2 p+8\pi T b=0 \, .
\end{align}
Now we can assume that Eq.~\eqref{eq:Min_Energy_Bubble} is balanced by two dominant terms. Assuming the first term to be negligible we find
\begin{align}
b=-\frac{2T}{p}.
\end{align}
Since $b$ is negative, it follows that we can not neglect the first term in the equation. On the other hand, assuming the second term to be negligible gives
\begin{align}\label{eq:b1}
b=\left(\frac{\hbar^2\pi}{8m_e T}\right)^{1/4}.
\end{align}
Substituting~\eqref{eq:b1} into~\eqref{eq:Energy_Bubble} we obtain
\begin{align}
E=\left[ \frac{\hbar^2\pi^2}{2}\left(\frac{\hbar^2\pi}{8T}\right)^{-1} +\frac{4\pi p}{3}\left(\frac{\hbar^2\pi}{8 T m_e}\right)^{1/4}+4\pi T \right] \left(\frac{\hbar^2\pi}{8 T m_e}\right)^{1/2} \, .
\end{align}
Motivated by the physics of the problem, we consider the limit $\delta\rightarrow 0$. In this regime, the second term is dominant, which is inconsistent with our initial assumption. Finally, if we assume the third term to be negligible, then
\begin{align}\label{eq:b2}
b=\left( \frac{\pi \hbar^2}{4m_e p}\right)^{1/5}.
\end{align}
Substituting~\eqref{eq:b2} into~\eqref{eq:Energy_Bubble} we obtain
\begin{align}
E=\left[ \frac{\hbar^2\pi^2}{2m_e^{1/5}} \left( \frac{\pi \hbar^2}{4 p}\right)^{-4/5}+\frac{4\pi p}{3m_e^{1/5}}\left( \frac{\pi \hbar^2}{4 p}\right)^{1/5}+4\pi T \right] \left( \frac{\pi \hbar^2 }{4 p m_e}\right)^{2/5} \, .
\end{align}
In the limit $\delta\rightarrow 0$ the third term is negligible, which leads to a self-consistent estimate. It follows that the radius of the bubble at non-zero pressure will be given by
\begin{align}
b \sim \left( \frac{\pi \hbar^2}{4m_e p}\right)^{1/5}.
\label{Eq:radius_b}
\end{align}
This provides an important length scale in the problem that dictates the size of the computational domain that will be needed in our simulation to resolve the relevant physical scales of interest.
%By considering the surface tension term to be negligible compared to the others it is possible to estimate the radius of the electron bubble in the limit of small values of $m_e$ being  
%\begin{equation}\label{eq:Radius_Bubble}
%b\simeq \left( \frac{\pi \hbar^2 M^2}{2m_e V_0 \rho^2 }\right)^{1/5}.
%\end{align}
%The proof this last result is shown in Appendix ().

%\section{Numerical scheme}

\section{Initial condition \label{App:IC}}
In order to find the correct initial condition for the electron in the ground state we first need to solve the Helmholtz equation
\begin{align}
\nabla^2 \phi+k^2\phi=0 \, ,
\end{align}
in a sphere of radius $\pi/k$. The spherically symmetric modes are given by 
\begin{align}
\phi_0(r,\theta,\varphi)=\left(\frac{2k^3}{\pi\epsilon^3}\right)^{1/2}\frac{\sin(kr)}{kr} \, , \,\,\,\, \qquad r<\pi/k \, ,
\end{align}
where $k$ represents the different eigenvalues that can be supported by the system. For the ground state with energy $E_q$ given by Eq.\ \eqref{Eq:KinEn_Sphere}, we find
\begin{align}
k^2=\frac{m_e E_q}{m_4 \mu}=\frac{\hbar^2 \pi^2}{2 m_4 \mu b^2}=\frac{\pi^2 a^2}{b^2}=0.0342 \, ,
\end{align}
to obtain
\begin{align}
4\pi\int _0^{\pi/k} |\phi|^2 r^2 dr=\frac{4\pi}{\epsilon^3} \, .
\end{align}
For the superfluid wave-function we choose a density profile given by 
\begin{align}
\psi_0(r,\theta,\varphi)&= \left\{ \begin{array}{cc}
\tanh\left(\frac{r-b}{\sqrt{2}}\right) \, , &  \,\,\,\,\,\,\,\, r\ge b \, , \\
0  \, , & \,\,\,\,\,\,\,\, r\le b \, .
\end{array} \right.
\end{align}
With the above initial conditions for the two fields $\psi(\bx,0)=\psi_0(\bx)$ and $\phi(\bx,0)=\phi_0(\bx)$, we can then integrate the system of equations
\begin{align}\label{Eq:system}
\frac{\partial}{\partial \eta}{\psi  \choose \phi }= \frac{1}{2}\begin{bmatrix}  \nabla^2 & 0 \\ 0 & \delta^{-1}\nabla^2 \end{bmatrix}{\psi  \choose \phi }-\frac{1}{2}\begin{bmatrix} \gamma |\phi|^2 + |\psi|^2 & 0 \\ 0 & \delta^{-1}\zeta^2 |\psi|^2 \end{bmatrix}{\psi  \choose \phi }, 
\end{align}
with respect to the imaginary time, $\eta$, until the system converges to the desired level of accuracy. This gradient flow method was applied until by Eq.~\eqref{eq:err} satisfied the threshold $\text{Err}<10^{-7}$ in Eq.~\eqref{Eq:system} (see below). 

%\subsection{Gradient flow method}
%
%In order to find the ground state of~\eqref{Eq:SCH_ad_1} we have to integrate the diffusive equation
%\begin{align}\label{imm_time}
%\frac{\partial \phi}{\partial \eta}=\frac{1}{2\delta}\nabla ^2\phi -\frac{1}{2\delta}\left(q^2|\psi(x,t)|^2\phi+\frac{xcE\delta}{\mu}\right)\phi
%\end{align} in term of the parameter $\eta$.
%We note that the equation~\eqref{imm_time} corresponds to solve equation~\eqref{SCH_ad} in the imaginary time direction.
%Starting with an initial condition $\phi(x,\eta_0)=\phi(x,t_0)$ we then use a SSSM
%\begin{align}
%\phi(x,\eta^*)= e^{(d\eta/2) N(x)}e^{d\eta L}e^{(d\eta/2) N(x)}\phi(x,\eta_0)
%\end{align}
%where $N(x)$ is defined as
%\begin{align}
%N(x)= -\frac{q^2|\psi(x,t_1)|^2+xcE\delta/\mu}{2\delta},
%\end{align}
%and at the end we evaluate the $\mathsf{L}^2$ norm given by
%\begin{align}
%\text{err}=\int |\phi(x,\eta^*)-\phi(x,\eta_0)|^2 dx.
%\end{align}
%Such method is iterated considering $\phi(x,\eta^*)$ as new initial condition, until the value of err results smaller than an arbitrary threshold.
%The presence of the $x$-dependent term $\frac{xcE\delta}{\mu}$ in the equation of motion of the bubble could in principle cause an issue with the choice of periodic boundary conditions.
%However, since the electron-bubble is a highly localized object the wave function will decay exponentially outside the confining cavity. Hence considering a sufficiently large domain allows us to decompose the wave function in the Fourier space.

\section{Numerical Integration of Equations of Motion \label{App:Num}}
For all our numerical simulations, we have assumed periodic boundary conditions that permit Fast Fourier Transforms (FFTs) to be used to evaluate the kinetic energy terms appearing in our system of equations. To advance our equations forward in (real or imaginary) time, we use a symmetric Strang splitting pseudo-spectral method for Eq.~\eqref{Eq:system}. This leads to
\begin{align}\label{Eq:Grad_Flow}
{\psi(\vv{x},\eta+\Delta \eta)  \choose \phi(\vv{x},\eta+\Delta \eta) }= e^{(\Delta\eta/2) \hat{\mathcal{N}}(\vv{x})}e^{\Delta\eta \hat{\mathcal{L}}}e^{(\Delta\eta/2) \hat{\mathcal{N}}(\vv{x})}{\psi_0 (\vv{x},\eta) \choose \phi_0(\vv{x},\eta) } \, .
\end{align}
In equation~\eqref{Eq:Grad_Flow}, $\hat{\mathcal{N}}(\vv{x})$ is defined in the physical space as
\begin{align}
\hat{\mathcal{N}}(\vv{x})=\begin{bmatrix}  \hat{\mathcal{N}}_{GP}& 0 \\ 0 & \hat{\mathcal{N}}_e\end{bmatrix}= -\frac{1}{2}\begin{bmatrix} \gamma |\phi|^2 + |\psi|^2 & 0 \\ 0 & \delta^{-1}\zeta^2 |\psi|^2 \end{bmatrix} \, ,
\end{align}
whereas $\hat{\mathcal{L}}(\vv{x})$ is defined in Fourier space as
\begin{align}
\hat{\mathcal{L}}(\vv{k}) = \int \hat{\mathcal{L}}(\vv{x}) \text{e}^{i \bk \cdot \bx} \d^3\bx  = \begin{bmatrix}  \hat{\mathcal{L}}_{GP}& 0 \\ 0 & \hat{\mathcal{L}}_e\end{bmatrix}=-\frac{1}{2}\begin{bmatrix}  |\vv{k}|^2 & 0 \\ 0 & \delta^{-1}|\vv{k}|^2  \end{bmatrix} \, .
\end{align}
This method is iterated until the $\mathsf{L}^2$ norm defined as
\begin{align}\label{eq:err}
\text{Err}=\int \left |{\psi(\vv{x},\eta+\Delta \eta)  \choose \phi(\vv{x},\eta+\Delta \eta) }-{\psi(\vv{x},\eta)  \choose \phi(\vv{x},\eta) }\right |^2 \d^3\vv{x} \, ,
\end{align}
drops below a specified threshold.
%\subsection{Dynamical evolution}\label{Sec:Dynamical evolution}
Once the equilibrium state of the system has been determined, we set $Q\neq 0$ and integrate Eq.~\eqref{Eq:GP_ad_1} to study the dynamics of the superfluid and electron bubble in the adiabatic approximation. 
%The numerical integration is performed using periodic boundary conditions and adopting the same pseudo-spectral method that is described in the previous section. 
The evolution of the superfluid from time $t_0$ to time $t=t_0+\Delta t$ is given by
\begin{align}
\psi(\vv{x},t+\Delta t )= e^{i(\Delta t/2)\hat{\mathcal{L}}_{GP} }e^{i\Delta t \hat{\mathcal{N}}_{GP}(\vv{x},t_1)}e^{i(\Delta t/2) \hat{\mathcal{L}}_{GP}}\psi(\vv{x},t) \, ,
\end{align} 
where $t_1=\Delta t/2+t$. We note that in contrast to Eq.~\eqref{Eq:Grad_Flow}, this choice of splitting allows us to evaluate the the ground state of the bubble once within each time step of the simulation.
The nonlinear operator $\hat{\mathcal{N}}_{GP}$ is defined in terms of 
\begin{align}
\psi(\vv{x},t_1)=e^{i( \Delta t/2) \hat{\mathcal{L}}_{GP}}\psi(\vv{x},t) \, ,
\end{align}
while $\phi(\vv{x},t_1)$ corresponds to the ground state for an electron governed by Eq.~\eqref{Eq:SCH_ad_1} in the presence of an external potential given by $({\zeta^2}/{2\delta})|\psi(\vv{x},t_1)|^2$.

In order to find the ground state of Eq.~\eqref{Eq:SCH_ad_1} we use the gradient flow method described above but applied only to the Schr$\ddot{\text{o}}$dinger equation now given by
\begin{align}\label{imm_time}
\frac{\partial \phi}{\partial \eta}=\frac{1}{2\delta}\nabla ^2\phi -\frac{1}{2\delta}\left(\zeta^2|\psi|^2\phi+yQ+E_e\right)\phi \, .
\end{align}
We note that 
%the equation~\eqref{imm_time} corresponds to solve equation~\eqref{SCH_ad} in the imaginary time direction.
%Starting with an initial condition $\phi(x,\eta_0)=\phi(x,t_0)$ we then use a SSSM
%\begin{align}
%\phi(x,\eta^*)= e^{(d\eta/2) N(x)}e^{d\eta L}e^{(d\eta/2) N(x)}\phi(x,\eta_0)
%\end{align}
%where $N(x)$\footnote{We note that for simplicity we used the same notation introduced in the split step method for the GP equation, however $N$ is now a linear operator.} is defined as
%\begin{align}
%N(x)= -\frac{q^2|\psi(x,t_1)|^2+xcE\delta/\mu}{2\delta},
%\end{align}
%and at the end we evaluate the $\mathsf{L}^2$ norm given by
%\begin{align}
%\text{err}=\int |\phi(x,\eta^*)-\phi(x,\eta_0)|^2 dx.
%\end{align}
%Such method is iterated considering $\phi(x,\eta^*)$ as new initial condition, until the value of err results smaller than an arbitrary threshold.
the presence of the $y$-dependent term, ${yQ\phi}/{2\delta}$, appears to be inconsistent with the use of  periodic boundary conditions along the $y$-coordinate direction. This difficulty is circumvented by noting that since the bubble is a localized object that is confined within the cavity created by the potential of the surrounding superfluid, the wave-function will decay exponentially outside this cavity. Indeed, we exploit this property of $\phi$ to allow us to solve the Schr\"{o}dinger equation on a truncated domain (see Fig.\ \ref{fig:chaos_large}). On the other hand, the motion of the electron bubble towards the boundaries can lead to numerical instabilities due to the discontinuous form of the potential arising from the last term in Eq.~\eqref{imm_time} across the boundaries. To avoid this, we apply a coordinate transformation that re-centers the bubble within the computational domain after a time interval $\Delta t_s$. The spatial translations are defined by setting  $\bx'=\bx-{\bf x}_{CM}(\Delta t_s)$, where
%one should adopt a frame of reference comoving with the bubble, to ensure the bubble staying in the centre of the computational domain. However, the bubble is moving at non-constant velocity, hence, additional terms in the equations of motion~\eqref{GP_ad} and ~\eqref{SCH_ad} will arise due to the choice of a non-inertial frame of reference. In practice, we can avoid this by letting the superfluid-bubble system to cover a time interval $\Delta t$ and thereafter performing a space-translation in the $y$-direction given by $y'=y-y_{CM}(\Delta t)$, where
\begin{align}\label{Eq:yCM}
{\bf x}_{CM}(\Delta t_s)=\frac{\int \bx |\phi(\vv{x},\Delta t_s)|^2 \d^3\vv{x}}{\int  |\phi(\vv{x},\Delta t_s)|^2 \d^3\vv{x}} \, ,
\end{align}
is the centre of mass of the bubble at time $\Delta t_s$. To keep track of the real position of the ion, we evaluate the cumulative displacement of the bubble by defining
\begin{align}
{\bf X}_{CM}(t_i)= {\bf X}_{CM}(t_i-\Delta t_s)+\sum_{j<i} \bx_{CM}(\Delta t_s)_j \, , 
\end{align}
where $i,j \ge1$. 
%{\color{red} Since the GP equation~\eqref{Eq:GP_ad_1} is invariant with respect to space translation we just need to express - not invariant if we include electric field potential? (I think this needs to be removed.)}
The condensate wave function $\psi(\vv{x}',t)$ in the new frame of reference can then be recovered from
\begin{align}
\psi(\vv{x}',t)=\hat{\mathcal{F}}^{-1}\left[e^{i {\bf k} \cdot {\bf x}_{CM}}\hat{\mathcal{F}}\left[\psi(\vv{x},t)\right]\right]
\end{align}
where $\hat{\mathcal{F}}$ stands for the fast Fourier transform. Using this newly evaluated wavefunction, computing $\phi$ can then be simply reduced to finding a new ground state subject to the shifted potential $|\psi(\vv{x}',t)|^2({\zeta^2}/{2\delta})$.\\

\section{Projected Gross-Pitaevskii equation \label{App:PGPE}}

An issue that arises when applying pseudo-spectral numerical methods applied to non-linear partial differential equations is the well known aliasing error that is caused from having a finite truncation in Fourier space\cite{NumericalRecipes}. To understand the source of the problem, we will express the GP equation (with $\gamma=0$) in terms of Fourier harmonics, such that
\begin{align}\label{eq:GP_fourier}
i\frac {d A_{\bk}}{dt}=\frac{k^2}{2}A_{\bk}+\frac{1}{2} \sum_{\bk_1,\bk_2}A_{\bk_1}A_{\bk_1+\bk_2}^*A_{\bk+\bk_2} \, ,
\end{align}
where
\begin{align}
\psi(\bx,t)=\sum_{\bk}A_{\bk}(t)e^{i \bk\cdot \bx} \, .
\end{align}
The essence of the aliasing error can now be understood by focussing on a periodic 1D system discretised  on $n_{max}$ collocation points in a domain of length $L_x$. This leads to $k=n\Delta k$ where $\lbrace n\in \mathbb{Z}: -n_{max}<n\le n_{max}\rbrace$ and $\Delta k=2\pi/L_x$. Therefore, the number of modes is defined up to a cut-off scale given by $k_{max}=n_{max}\Delta k/2$. We note that for such a discrete system, the harmonic $e^{i nx\Delta k}$ is equivalent to $e^{i (n+jn_{max})x\Delta k} \, \forall \, {j\in\mathbb{Z}}$. In general, the non-linear term can excite modes with a higher harmonic (e.g. the interaction of the modes corresponding to $k_1$ and $k_2$ and lying within the range $-k_{max}<k_1,k_2\le k_{max}$, can excite a $k_1+k_2$ mode). It follows that if not accounted for correctly, this $k_1+k_2$ mode will project back onto the modes within the range $-k_{max}<k_1,k_2\le k_{max}$ leading to inaccurate solution of the equations. This is the essence of the aliasing phenomena. 

To avoid such errors that result in the biasing of the amplitude of the lower modes, we introduce a low-pass filter acting in Fourier space. Such a filter consists of truncating all the modes higher than $2k_{max}/3$. To apply such a filter, we define a projector $\hat{\mathcal{P}}$ acting on the Fourier space as
\begin{align}
\hat{\mathcal{P}}[A_{k}]=\Theta(2k_{max}/3-|k|)A_{k},
\end{align}
where $\Theta(\cdot)$ is the Heaviside step function. Generalising these arguments to 3D leads to the truncated form of the GP equation (TGP):
\begin{align}
i\frac{\partial \psi}{\partial t}=\hat{\mathcal{P}}\left[-\frac{1}{2}\nabla^2\psi+\frac{1}{2}\hat{\mathcal{P}}[|\psi|^2]\psi\right] \, .
\end{align}
This equation can be derived from the truncated Hamiltonian 
\begin{align}
H=\int \left(\frac{1}{2}\hat{\mathcal{P}}\left[ |\nabla\psi|^2\right] +\frac{1}{4}\left(\hat{\mathcal{P}}\left[|\psi|^2\right]\right)^2\right)\d^3\vv{x} \, .
\end{align}
As shown by Krstulovic and Brachet \cite{Krstulovic2011}, such a system also conserves the number of particles and the linear momentum. From these considerations, it follows that if we include the interaction with the electron wave function, we can finally write the projected Gross-Clark equation for the superfluid, as
\begin{align}\label{eq:TGG}
i \frac{\partial \psi}{\partial t} = \hat{\mathcal{P}}\left[-\frac{1}{2}\nabla^2 \psi+\frac{\gamma}{2}\hat{\mathcal{P}}\left[|\phi_g|^2\right]\psi +\frac{1}{2}\hat{\mathcal{P}}\left[|\psi|^2\right]\psi\right] \, .
\end{align}
We have found that, in practice, introducing this projector helps stabilise our numerical scheme.\\
%{\color{red}

\section{Experimental data of drift velocities.\label{App:Expt}}
\begin{figure}[h]
 \centering
      \includegraphics[scale=0.4]{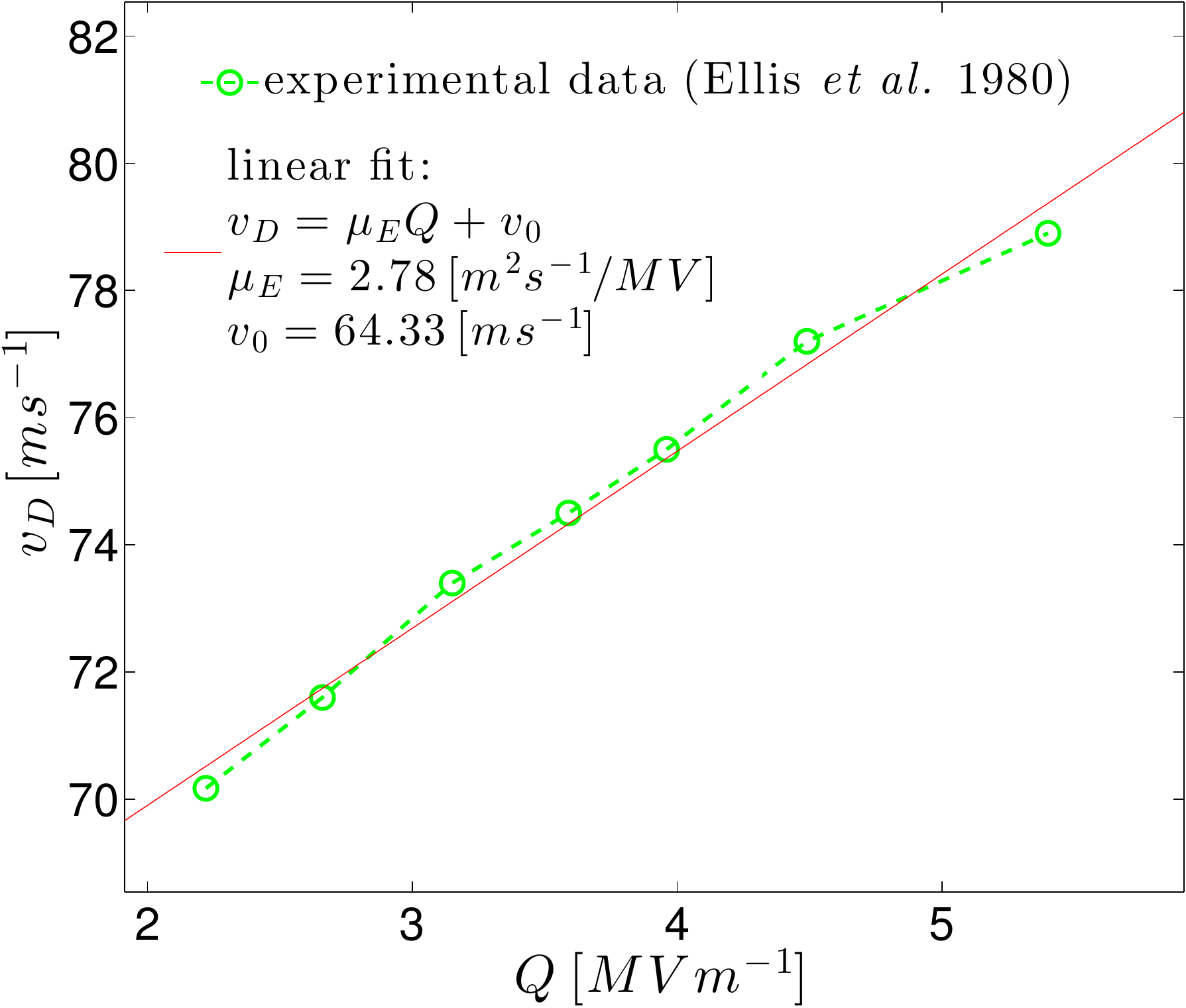}
 \caption{Plot of the measured drift velocity of a negative ion taken from [\onlinecite{Ellis1980}] under different applied electric fields. A linear relation between $v_D$ and $Q$ is found by fitting the experimental data with a first order polynomial $v_D=\mu_E Q+v_0$. \label{fig:Ellis}}
\end{figure}

Here we present the experimentally measured drift velocities of an ion moving, at pressure $p=2.5$ MPa and temperature $T=0.34$ K, under different values of the electric field $Q$. The data is taken from Ellis {\it et al.} \cite{Ellis1980}.
In particular, we focus on the range $v_D(Q)> 70\, \text{m}\text{s}^{-1}$ where, according to [\onlinecite{Ellis1980}], the drift velocities does not follow the expected trend predicted from assuming that the main source of drag acting on the motion of the ion is related to the emission of roton-pairs. In Fig.~\ref{fig:Ellis}, we have plotted the data for $v_D$ as a function of $Q$. As can be seen, a linear relation exists over the considered range of $Q$ which provides a measured value of the mobility equal to $\mu_{E}=2.76 \,\text{m}^2\text{s}^{-1}\text{M}^{-1}\text{V}^{-1}$.
%}

%\bibliographystyle{aipauth4-1}
\bibliographystyle{aipnum4-1}
\bibliography{PRB}

\end{document}